\documentclass[fleqn,usenatbib]{mnras}

\usepackage{newtxtext,newtxmath}

\usepackage[T1]{fontenc}
\usepackage{ae,aecompl}

\usepackage[normalem]{ulem}
\usepackage{hyperref}
\usepackage{subfig}
\usepackage{booktabs}
\usepackage{supertabular}

\usepackage{graphicx}	
\usepackage{amsmath}	
\usepackage{amssymb}	
\usepackage{soul}
\newcommand{\angstrom}{\mbox{\normalfont\AA}}
\newcommand\specfigwidth{1.8}
\newcommand\tsnefigwidth{1.0}

\title[Outliers and similarity in APOGEE]{Detecting outliers and learning complex structures with large spectroscopic surveys - a case study with APOGEE stars}

\author[I. Reis et al.]{
Itamar Reis$^{1}$\thanks{E-mail: itamarreis@mail.tau.ac.il},
Dovi Poznanski$^{1}$,
Dalya Baron$^{1}$,
Gail Zasowski$^{2}$,
and Sahar Shahaf$^{1}$
\\
$^{1}$School of Physics and Astronomy, Tel-Aviv University, Tel-Aviv, 69978, Israel\\
$^{2}$Department of Physics and Astronomy, University of Utah, Salt Lake City, UT 84112, USA\\
}

\date{Accepted XXX. Received YYY; in original form ZZZ}

\pubyear{2017}

\begin{document}
\label{firstpage}
\pagerange{\pageref{firstpage}--\pageref{lastpage}}
\maketitle

\begin{abstract}
In this work we apply and expand on a recently introduced outlier detection algorithm that is based on an unsupervised random forest. We use the algorithm to calculate a similarity measure for stellar spectra from the Apache Point Observatory Galactic Evolution Experiment (APOGEE). We show that the similarity measure traces non-trivial physical properties and contains information about complex structures in the data. We use it for visualization and clustering of the dataset, and discuss its ability to find groups of highly similar objects, including spectroscopic twins. Using the similarity matrix to search the dataset for objects allows us to find objects that are impossible to find using their best fitting model parameters. This includes extreme objects for which the models fail, and rare objects that are outside the scope of the model. We use the similarity measure to detect outliers in the dataset, and find a number of previously unknown Be-type stars, spectroscopic binaries, carbon rich stars, young stars, and a few that we cannot interpret. Our work further demonstrates the potential for scientific discovery when combining machine learning methods with modern survey data.
\end{abstract}






\section{Introduction}

Extracting and analyzing information from ongoing and future astronomical surveys, with their increasing size and complexity, requires astronomers to take advantage of the tools developed in the (also rapidly growing) fields of data science and machine learning. Most commonly in astronomy, these methods enable detection or classification of specified objects using supervised machine learning (ML) algorithms, while unsupervised ML is used to search for correlations or clusters in high dimensional data. Recent examples for such work are 
\citet{bloom12} - identification and classification of transits and variable stars using imaging, \citet{meusinger12} - outlier detection with quasar spectra, \citet{masci14} - classification of periodic variable stars using photometric time-series, \citet{baron15} - clustering diffuse interstellar band lines based on their pairwise correlation,  \citet{miller17} - Star-Galaxy classification based on imaging.
 A review of data science applications in astronomy can be found in \citet{ball10}.

In this work we focus on unsupervised exploration of a dataset based on a similarity matrix, containing a pair-wise similarity measure between every two objects (the simplest possible measure being the euclidean distance between the features of two objects). We show that such a similarity matrix (or its inverse, the distance matrix) is a powerful tool for exploring a dataset in a data-driven way. We calculate an unsupervised Random Forest based similarity measure for stellar spectra and show that, without any additional input other than the spectra themselves, the similarity matrix traces physical properties such as metallicity, effective temperature and surface gravity. This allows us to visualize the complex structure of the dataset, to query for similar objects based on their spectra alone, to put an object in the context of the general population, to scan the dataset for different object types, and to detect outliers.

These possibilities are only partly available with traditional representation of an object in an astronomical database, i.e. by its fit parameters. Fitting a model requires making assumptions about the object. This can work well for a large fraction of the data, but usually cannot account for all the objects nor all of the features. In datasets composed of astronomical spectra, the model fitting is usually based on spectral templates, that do not cover the entire range of parameters available in the dataset. Furthermore, templates are usually not available for rare or unexpected objects.  This leaves a fraction of the objects, even if well understood, not well fitted, and impossible to query using the database.

Generative models have recently gained popularity within the astronomical community and outside of it, as they solve some of the issues raised above. Generative models are generated from the dataset, with few to no assumption about the data structure and distribution of information content. These models, which are purely data-driven, have been shown to generalise well, and describe even the most extreme objects in the sample without the need for dedicated treatment. An example of generative models is generative adversarial neural networks (GANs), for a recent use in astrophysics see \citet{schawinski17}. In this work, we show that the unsupervised RF algorithm can be viewed as a generative model, as it grasps complex features in the dataset, and is able to describe the most extreme objects in the sample in the same context as the common ones.

Perhaps the most intriguing usage of a similarity matrix is outlier detection. Outliers in a dataset can have different origins and interpretations. Some are measurement or data processing errors, and others are objects not expected to be in the dataset, extreme and rare objects, and most importantly, unknown unknowns - objects we did not know we should be looking for. In addition, in astronomy, rare objects could actually be important and common evolutionary phases that are short lived, and therefore challenging to observe.  It is worth noting that finding the mundane outliers is still useful in order to clean the dataset from erroneous and unwanted objects, to allow for a better analysis of the rest of the sample.  

Outlier detection algorithms can be divided into different types: (i) Distance based algorithms, which we use in this work, relying on a (case specific) definition of a pair-wise distance between the objects, (ii) Probabilistic algorithms, based on estimating the probability density function of the data, (iii) Domain based algorithms, which create boundaries in feature space, (iv) Reconstruction based algorithms, which model the data and calculate the reconstruction error as a measure for novelty, and (v) Information-theoretic algorithms, which use the information content of the data (for example by computing the entropy of the data), and measure how specific objects in the dataset change this value. For a review see \citet{pimentel14}. We use a distance based algorithm as it allows us to explore the data in additional ways, as discussed above. One thing to note is that for a large and complex enough dataset it is likely that there is not a single outlier detection algorithm that is best, i.e. one algorithm that detects all the interesting outliers. In general, different algorithms could be sensitive to different types of outliers. An obvious test for such an algorithm is whether it detects the expected outliers, if it does then it could be worthwhile to investigate all the detected outliers. But even then there is no guarantee that a different algorithm would not detect additional interesting objects.


In this work we expand the outlier detection algorithm presented in \cite{baron17a} \footnote{Code can be found at https://github.com/dalya/WeirdestGalaxies} and apply it to infrared stellar spectra. The core of the algorithm is calculating a distance matrix of the objects in the sample This distance is based on Random Forest Dissimilarity. For Random Forest (RF) see \citet{breiman84,breiman01}, for RF Dissimilarity see \citet{breiman03,shi06}. There are many possible choices for a similarity measure, a simple example being the euclidian distance between the features of the objects. See \citet{yang06} for a survey of distance metric learning. It is known (see \citet{yang06}  and references therein) that a good choice of distance metric can improve the accuracy of K-nearest-neighbor classification (a common application of a distance metric), over simple euclidian distances.  Similarly to outlier detection algorithms, there is no best distance metric, even for a specific dataset. As there are many possible usages for a distance metric, it is even less clear how such best distance metric would be defined. An intuitive reason to use RF dissimilarity is that, as described below, it is sensitive to the correlation between different features. This is is often of importance in spectra. For instance line ratios are usually of more interest then the strength of a single line. A euclidian distance metric will be more sensitive to strength of single lines. See   \citet{garcia-dias18} for an application of an euclidian distance metric in a clustering algorithm with APOGEE spectra.

\cite{baron17a} applied this algorithm to find outliers in galaxy spectra from Sloan Digital Sky Survey \citep[SDSS; ][]{eisenstein11} and used the distance matrix to detect outliers. They found spectra showing various rare phenomena such as supernovae, galaxy-galaxy gravitational lenses, and double peaked emission-lines, as well as the first reported evidence for AGN-driven outflows, traced by ionized gas, in post starburst E+A galaxies. The last discovery is discussed in \citet{baron17}. The algorithm was applied to galaxy spectra using the flux values at every wavelength as features for the RF (i.e., without generating user defined features). Here we do the same with stellar spectra from the Apache Point Observatory Galactic Evolution Experiment \citep[APOGEE, ][]{majewski16}, which is part of the SDSS-III, and explore additional applications of the distance matrix produced by the algorithm. We visualize the distance matrix using the t-Distributed Stochastic Neighbor Embedding (t-SNE) algorithm \citep{maaten08}, find objects which are similar to objects of interest, and find the most similar objects in the dataset (that is - spectroscopic twins).

This paper is organized as follows. Section \ref{sec:apogee_data} describes the APOGEE dataset we use in this work. In section \ref{sec:tsne} we use t-SNE to visualize the distance matrix produced by our algorithm, and show that it traces stellar parameters. We use the distance matrix to find groups of similar objects, and spectroscopic twins.  In section \ref{nnws_subsec} we discuss ways to select and classify outliers efficiently. In section \ref{resultssec} we present the classification of the outliers we detected. We summarize in section \ref{sec:sum}.

\section{APOGEE spectra}
\label{sec:apogee_data}
The 14th SDSS data release \citep[DR14;][]{abolfathi17} contains the first data release for the APOGEE-2 survey. The APOGEE-2 survey consists of high resolution ($\mathrm{R} \sim$ 22,500), high signal to noise ratio (typically $S/N > 100$), infrared $H$-band (1.51-1.70 $\mu$m) spectra for $\sim$ 263,000 different stars. The APOGEE-2 main survey spans all galactic environments (bulge, disk, and halo) and is composed mainly of red giant stars. The main survey targets were chosen using a cut on the $H$-band magnitude, gravity-sensitive optical photometry, and dereddened $\left( J - K_{s} \right)_0$ color limits. The color limit and optical photometry criteria are intended to separate red giants from main sequence dwarfs.  The APOGEE-2 dataset contains  $\sim$ 32,000 non main survey targets, including $\sim$ 13,300 ancillary targets, and $\sim$ 27,000 hot stars used for telluric correction. More details on the target selection in APOGEE-2 are in \citet{zasowski17}. A large fraction of the work done with APOGEE data is devoted to investigating the Milky Way structure and evolution using chemical abundances and radial velocities (RVs) derived from the spectra; for examples see \citet{frinchaboy13,nidever14,bovy14,ness15,chiappini15,hayden15}. We note that APOGEE spectra are rich with information, and a single spectrum can contain hundreds of absorption lines.

The input to our algorithm is the pseudo-continuum normalized (PCN) spectrum. The pseudo-continuum normalization procedure is done with the APOGEE Stellar Parameters and Chemical Abundances Pipeline \citep[ASPCAP;][]{garcia-perez16} in order to remove variations of spectral shape arising from interstellar reddening, errors in relative fluxing, detector response, and broad band atmospheric absorption. The APOGEE spectra contain two gaps in wavelength. Our preprocessing stage consists of removing flux values in these gaps (these values are set to zero in the original PCN spectrum), as well as interpolating the spectra to the same wavelength grid. This leaves us with 7,514 flux values per object, which are the features used by the outlier detection algorithm.

Applying our algorithm to the APOGEE-2 spectra from DR14, it became clear that many objects have faulty PCN spectra (these objects are discussed in section \ref{resultssec}). Our algorithm naturally classifies these objects as outliers, making it harder to find the more interesting outliers. For this reason, since DR13 does not suffer from this contamination, we apply the algorithm to DR13 data as well. DR13 contains spectra for 163,000 stars, 25,000 of which are non main survey. DR13 contains results from APOGEE-1, for which the target selection is somewhat different, and is described in \citet{zasowski13}. Unless otherwise stated the results presented in this paper refer to DR14, APOGEE-2 (which we will refer to as APOGEE) data.

We use only objects with $S/N > 100$, of which there are 193,556 in DR14 (107,390 in DR13). The input data size is therefore the product of the number of objects by the number of features (wavelengths). The reason for not using low $S/N$ objects is that when included, objects with spectra dominated by noise are detected as outliers. We note that for the high $S/N$ objects we used, the weirdness score and the $S/N$ are not correlated.


\section{Exploring the APOGEE dataset using a distance matrix}
\label{sec:tsne}

Using our distance matrix to find physically interesting outliers and study the structure of the dataset requires it to retain the complex information that we see in each object in the sample, which is a non-trivial task. In this section we explore what type of information our distance matrix contains. \cite{baron17a} have seen some hints that the RF distance matrix contains a wealth of complex spectral information aggregated to a single number, the pair-wise distance, here to explore that question using visualization and dimensionality reduction tools.

\subsection{Random Forest Dissimilarity}
Briefly, the distance is calculated by the following procedure. First, synthetic data are created with the same marginal distributions as the original data in every feature, but stripped of the correlation between different features (the features in our application are the flux values at each wavelength of the spectra, as described below). Having two types of objects, one real and one synthetic, an RF classifier is trained to separate between the two. In the process of separating the synthetic objects with un-correlated features from the real ones, the RF learns to recognize correlations in the spectra of real objects. The RF is composed of a large number of classification trees, each tree is trained to separate real and synthetic objects using a subset of the data (the 'Random' in 'Random Forest' is referring to the randomness in which a subset of the data is selected for each tree, see \citet{breiman84,breiman01} for details). Having a large number of trees, the similarity $S$ between two objects (objects in the original dataset, i.e. real objects) is then calculated by counting the number of trees in which the two objects ended up on the same leaf (a leaf being a tree node with no children nodes), and dividing by the number of trees. This is done only for the trees in which both objects are classified as real. 
We define the distance matrix to be $D = 1 - S$. Using the distance matrix we can calculate a 'weirdness score' for every object, defined to be the average distance to all other objects. Below we refer to this weirdness score as $W_{all}$. See \citet{baron17a} for a detailed description of the algorithm. 

To build the distance matrix we use the scikit-learn implementation of Random Forest. The number of trees we used is 5000. We note that this number was necessary to reach convergence, i.e. increasing this number further does not alter the results. Every 200 trees are built using a random subset of 10000 objects.

\subsection{The t-SNE algorithm} 
t-SNE is a dimensionality reduction algorithm that is particularly well suited for the visualization of high-dimensional datasets. We use t-SNE to visualize our distance matrix. A-priori, these distances could define a space with almost as many dimensions as objects, i.e., tens of thousand of dimensions. Obviously, since many stars are quite similar, and their spectra are defined by a few physical parameters, the minimal spanning space might be smaller. By using t-SNE we can examine the structure of our sample projected into 2D.  We use our distance matrix as input to the t-SNE algorithm and in return get a 2D map of the objects in our dataset. In this map, nearby objects have a small pair-wise distance, and distant objects have a large pair-wise distance. The two t-SNE dimensions have no physical interpretation. Since the dimensionality in greatly reduced in the process, this is approximate, and breaks for large distances. That is, the map does not show the relative pair-wise distance between "far away" and "very far away" objects. The map does preserve small scale structure. 

The general idea of the t-SNE algorithm is quite simple - trying to preserve the distances of each object to its nearest neighbors (the number of which is determined by the \textit{perplexity} parameter), while forcing the distances to reside on a lower dimensional plane, in our case 2D. There is usually no single best t-SNE map. Maps calculated with different numbers of nearest neighbors can provide the user with different information about the dataset. For example a map calculated with 10000 nearest neighbors is not likely to show a cluster that contains 100 objects, while a map with 100 nearest neighbors is.  Other free parameters in t-SNE are of computational nature, and control speed vs. accuracy (accuracy of approximations done in different calculations inside the algorithm). A bad choice of parameters is usually manifested by a large fraction of the objects distributed randomly on the map. We consider a map in which all or almost all of the objects are located in structures to be a good map. Once we have that we can change the \textit{perplexity} to determine the 'scale' in which the objects are clustered. A guide for effective use of t-SNE is available in \citet{wattenberg2016how}.

We use the scikit-learn \citep[][]{pedregosa11} implementation of t-SNE. We note that to get informative maps we had to significantly increase the learning rate parameter (in the t-SNE map shown below it was set to 40,000) from its default value of 1000. The perplexity we used was 2000. Both of these parameters required adjustment when changing the number of objects in the distance matrix. Building the map took about 3 days of computation on a machine with 32 cores and 1TB of RAM. When using the current version of scikit-learn (0.17), t-SNE is using memory of about 8 times the size of the distance matrix. The memory usage will be significantly reduced in future scikit-learn versions. We used the development version of t-SNE which will be included in scikit-learn 0.19. With this version the memory usage was reduced by roughly a factor of 4, depending on the perplexity.

\subsection{A t-SNE map of the APOGEE dataset}
We apply the t-SNE algorithm to our RF dissimilarity distance matrix. The map produced can be especially informative when using different object attributes to color the points. 

\begin{figure*}

\subfloat[]{\includegraphics[width=\tsnefigwidth\columnwidth]{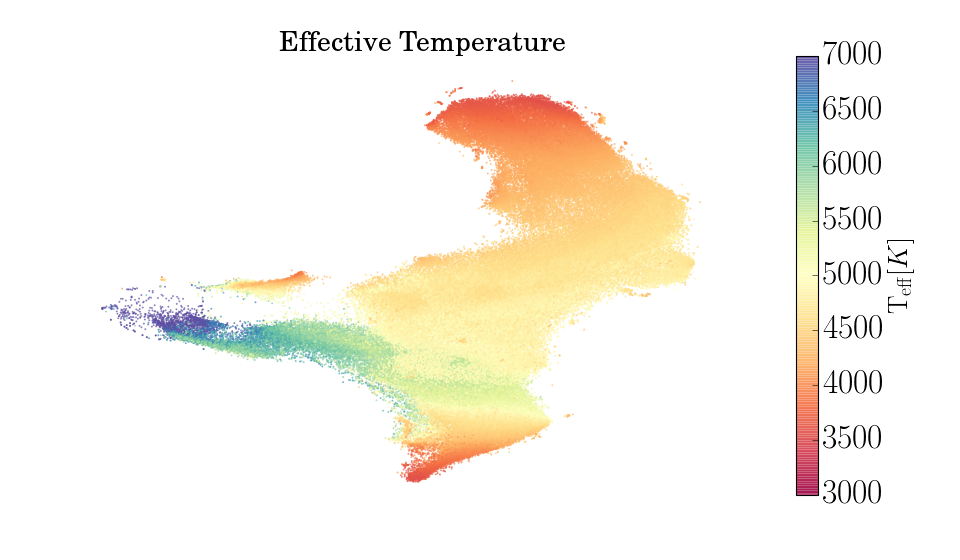} \label{subfig:tsne_teff}}
\subfloat[]{\includegraphics[width=\tsnefigwidth\columnwidth]{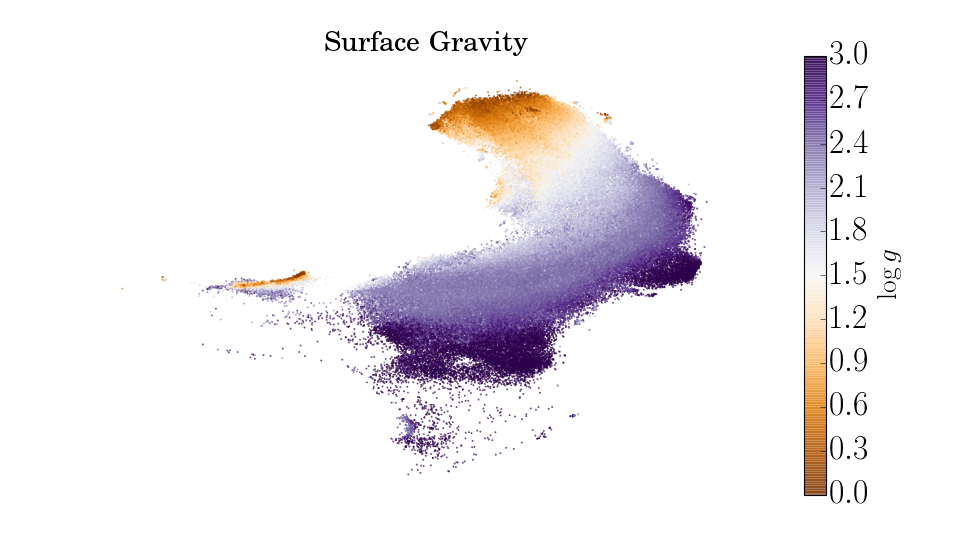}} \\[6pt]
\subfloat[]{\includegraphics[width=\tsnefigwidth\columnwidth]{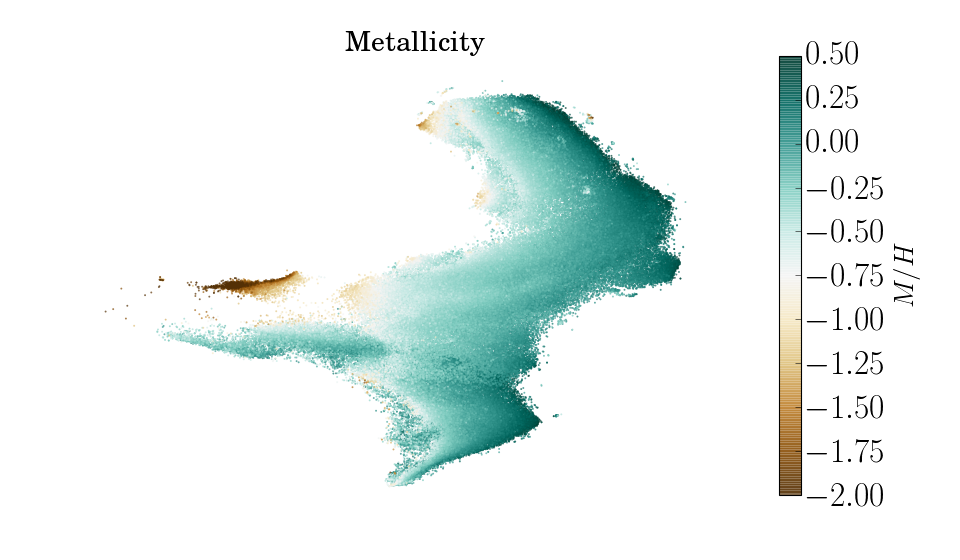}} 
\subfloat[]{\includegraphics[width=\tsnefigwidth\columnwidth]{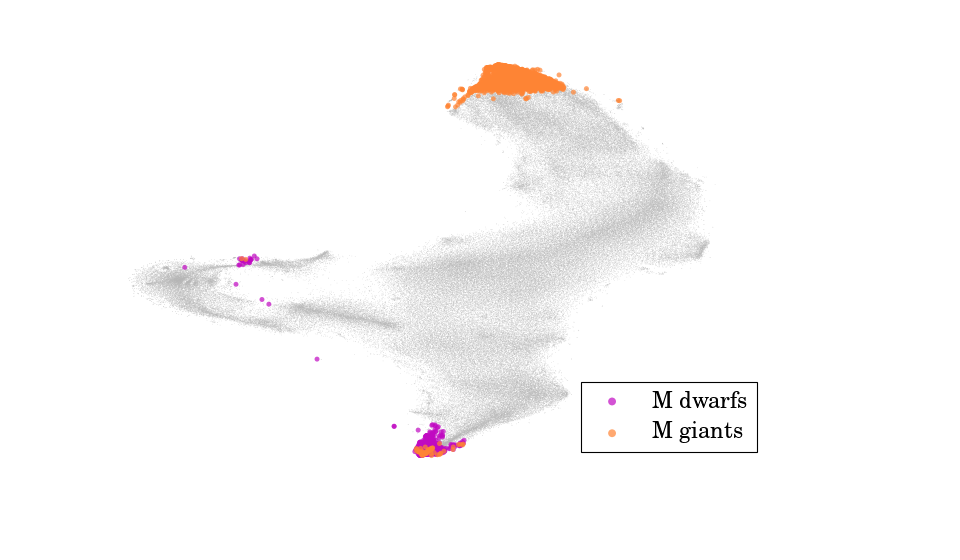}\label{subfig:tsne_m}} \\[6pt]
\subfloat[]{\includegraphics[width=\tsnefigwidth\columnwidth]{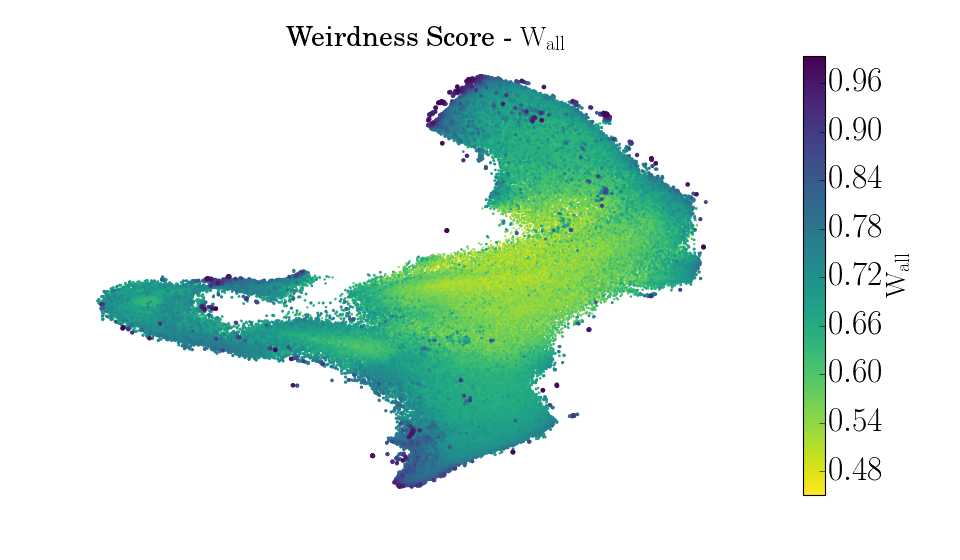}\label{subfig:tsne_wall}} 
\subfloat[]{\includegraphics[width=\tsnefigwidth\columnwidth]{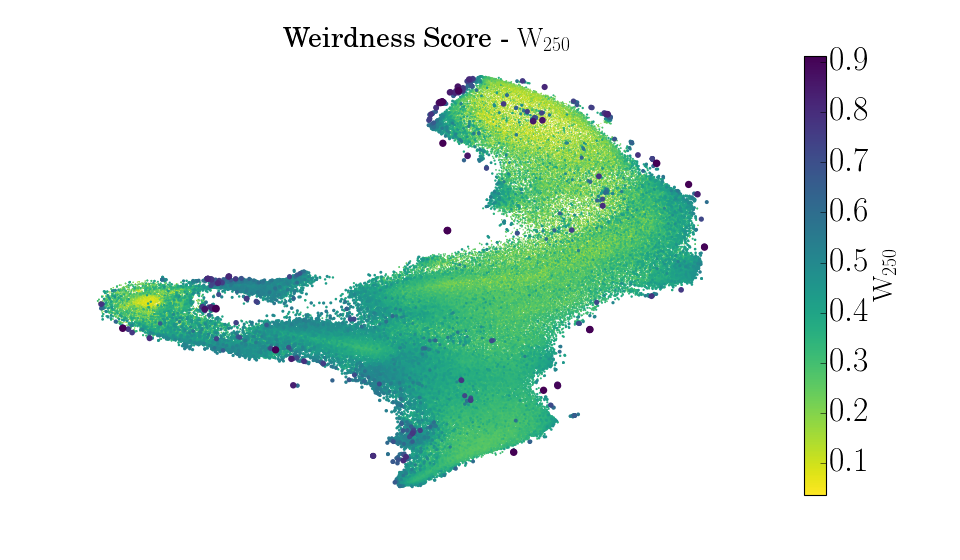}\label{subfig:tsne_w250}} \\[6pt]

\caption{t-SNE map of our distance matrix. Each point on the map represents a star, where spectrally similar objects cluster on small scales. The axes do not have any physical significance. In the different panels, different coloring schemes are presented. Panel (a): effective temperature, panel (b): surface gravity, panel (c): metallicity, panel (d): highlighted M-type stars. The values used for the different coloring are taken from ASPCAP. Stars with no available value for a parameter do not appear on the map. For example, many dwarf stars do not have $\log{g}$ values, so the clusters containing dwarfs disappear from the $\log{g}$ map. The complex structure of the sample is apparent. In panels (e) and (f) we color the map by the weirdness score. $W_{all}$ is in panel (e), and $W_{250}$ is in panel (f). We see that when using $W_{250}$, low $\mathrm{T}_{\mathrm{eff}}$ stars no longer dominate the high weirdness score population, and we get a more diverse outlier population that is spread on the t-SNE map.}
\label{fig:t_tsne}
\end{figure*}

In Figure \ref{fig:t_tsne} we use the following for color:
$\mathrm{T}_{\mathrm{eff}}$, highlight of M-type stars, metallicity, and $\mathrm{log} \left(g\right)$, based on the ASPCAP fit. Most of the objects lie in a right hand, mainly vertical, component of the map. In this part we see that the stars are sequenced by their surface gravity, where giants are located at the top and dwarfs at the bottom, as well as their effective temperature for which we get two separate sequences, one for dwarfs at the bottom of the map and  one for giants at the top of the map. We also see an horizontal sequence that follows the metallicity, high metallicity on the right. On the left hand side of the map we have the hotter stars in the APOGEE sample, including the stars used for telluric calibration. The very low metallicity stars are located near these telluric objects, both having mainly featureless spectra.

In panel \ref{subfig:tsne_m} we see that some M-type stars are located far from the rest. We manually inspect these objects as an example to see if this is due to the algorithm mis-locating a few objects, or if these objects are really different from the rest of their respective groups. We find that in this case the objects really have different looking spectra, with poor ASPCAP fitting. For example, some of these misplaced-M-type stars turn out to be B-type emission line stars (Be stars).

From the t-SNE maps we learn that our distance matrix is capable of aggregating non-trivial information about the objects in the sample. Figure \ref{fig:t_tsne} shows that the distance matrix holds information about various physical properties, namely the figure is showing sequences in the effective temperature, surface gravity, and metallicity. These properties, in addition to the chemical abundances, affect the spectral features in non-trivial and partly degenerate ways, which we see are captured in the distance matrix. 

The APOGEE pipeline derives the stellar parameters by means of best fitting templates. We see that some of the stars in the sample do not have derived parameter values. This is usually because these stars are extreme, at least with respect to the rest of the sample, and their stellar parameters fall outside the grid of spectral templates used by the pipeline. One can use the distance matrix to find these objects. As seen on the t-SNE map, the algorithm places the extreme objects next to the less extreme ones in a continuous sequence. In this sense we say that the similarity measure could be viewed as a generative model of the data. 

Seeing here that globally the distance matrix captures the structure of the dataset, in the next two subsections we see that it could also be used to investigate the dataset at a 'smaller scale' - by looking at the most similar objects.

\subsection{Object retrieval}
\label{sec:tsne_carbon}
\begin{figure}
\begin{center}
  \includegraphics[width=\columnwidth]{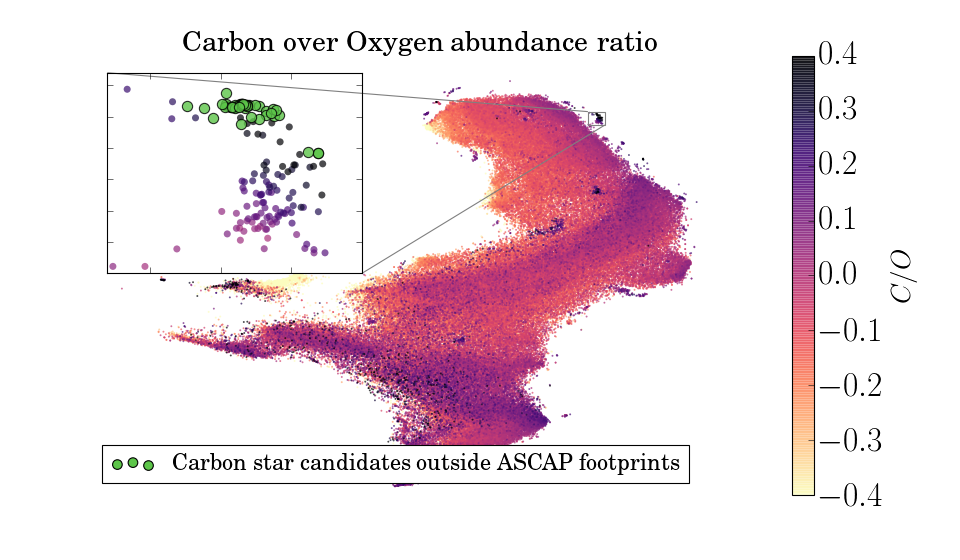}
  \caption{A t-SNE map colored by the carbon to oxygen abundance ratio, from ASPCAP. We focus on a cluster of carbon rich stars. We see that, according to a sequence visibly detectable on the map, objects at the high end of the sequence are not fitted by ASPCAP. Detecting these objects is possible using the similarity matrix.}
  \label{fig:tsne_carbon}
  \end{center}
\end{figure}

We can use the distance matrix to query the dataset for similar objects based on their spectra alone. We use the example of carbon rich stars to show that the algorithm can find objects that were not possible to find using their ASPCAP fit parameters. Carbon-rich stars have atmospheres with over-abundance of carbon compared to oxygen. In this case the excess carbon (i.e., carbon that is not tied in $\mathrm{CO}$) will allow $\mathrm{CN}$ (and other carbon molecules) to form. In regular stars there will be excess oxygen, that will form $\mathrm{OH}$. In Figure \ref{fig:tsne_carbon} we show a t-SNE map colored by the carbon to oxygen abundance ratio, from ASPCAP.

Focusing on a cluster of carbon rich stars we see that the objects are sequenced on the map according the the C/O abundance ratio. Most importantly we see that a number of objects without pipeline abundance value are located at the top of the sequence. We suggest that the pipeline is not able to fit these objects due to them having extreme abundances, and due to the difficulty of abundance analysis of spectra with very strong molecular lines. We manually inspect the 51 objects with no ASPCAP value shown in Figure \ref{fig:tsne_carbon} and see that they all show strong $\mathrm{CN}$ and weak $\mathrm{OH}$, typical for carbon rich stars. Moreover, all of the 35 objects that have SIMBAD \citep[][]{wenger00} entry are classified as carbon stars, making the 15 objects without SIMBAD entry carbon star candidates. We note that many of these objects were observed as part of the APOGEE-2 AGB stars ancillary program, see \citet{zasowski17}.

This in an example of a query for objects based on their spectra, instead of on their fit parameters. This is of importance for objects with bad or nonexistent fit parameters, that would otherwise be lost in the dataset. In a large enough dataset it is very likely such objects will exist,  these can be extreme cases of known phenomenon outside the range of the model,  rare objects outside the scope of the model, or other types of outliers in the dataset.

\subsection{Spectroscopic twins}
\label{sec:twins}
The distance matrix produced by the algorithm can be used for finding objects with spectra similar to each other, objects sometimes referred to as spectroscopic twins. This is trivially achieved by sorting the distance matrix.

One use of spectroscopic twins is measuring distances \citep{jofre15}. Twin stars will have the same luminosity, and if we know the distance to one of the stars (e.g. using parallax), we can calculate the distance to the other by comparing observed magnitudes. \citet{jofre15} looked for spectroscopic twins among 536 FGK stars, and detected 175 pairs with spectra indistinguishable within the errors.  As we work with a rather homogeneous sample of $\sim 10^5$ stars, we expect a large fraction of (multiple) twins. 

Example spectra for spectroscopic twins are shown in \ref{fig:specs_twins}. It can be seen that the pairs have virtually identical spectra. Note that in the top example in Figure \ref{fig:specs_twins} one of the spectra lacks an ASPCAP fit, preventing identification of a twin via these parameters. The middle pair have very similar parameters, while the bottom twins show more significantly different parameters. Our method finds them all, irrespectively.

One thing we note is that this method for finding twins works well for the common object types (common in terms of representation in the dataset), but might be less so for underrepresented types of objects (in which case we still get similar spectra, but not identical). The reason being that our unsupervised RF uses more extensively the features that are important for regular objects, and as a result it can separate objects based on subtle differences in these features. For other features that might not be important to most of the objects (for example hydrogen absorption or emission), the RF uses cruder cuts to separate objects. 

\begin{figure}
\begin{center}
  \includegraphics[width=\columnwidth]{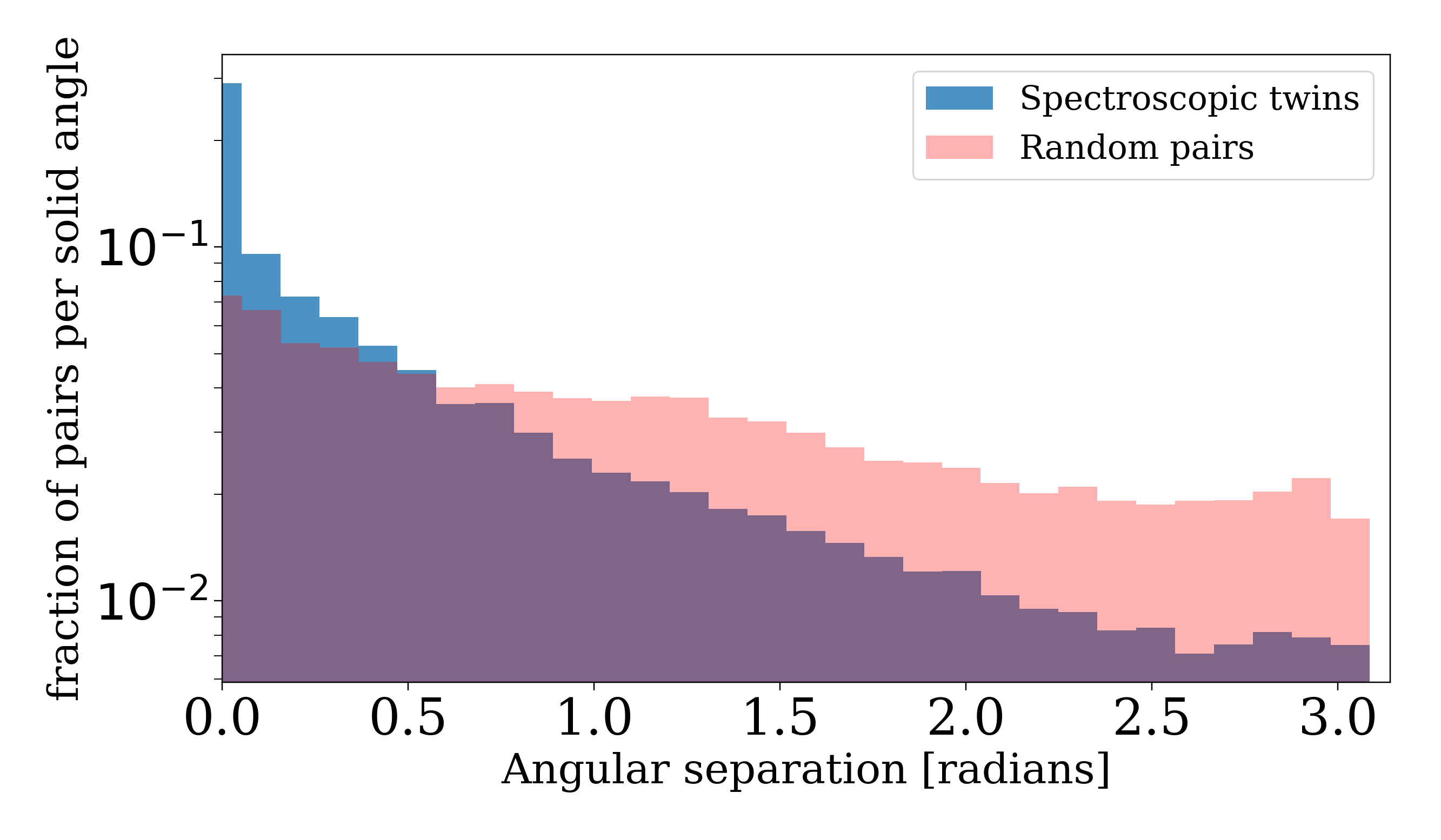}
  \caption{A comparison between angular separation of pairs of stars detected as having similar spectra, and random pairs of stars. The similar spectra stars shown here are each star in our sample and its nearest neighbor according to our distance matrix.}
  \label{fig:twins_sep}
  \end{center}
\end{figure}

As a test for the selection of twin objects we look at the angular separation of stars with similar spectra. APOGEE stars living in the same environment are more likely to have similar physical properties, and thus similar spectra. We expect that pairs of stars we detect as having similar spectra will have higher probability to be located near each other compared with random pairs of stars. We note that there is an observational effect playing a role here - the APOGEE sample is not uniform in different parts of the galaxy (for example,  dwarfs from the galactic bulge are too faint for APOGEE and are not observed). In Figure \ref{fig:twins_sep} we show the distribution of the angular separation of each star in our sample with its nearest neighbor (as defined by our distance matrix), compared to random pairs of stars. Clearly the twins we find are physically associated. We release our entire distance matrix, and we will update it with future data releases, allowing others to use the twins for further study. We note that our methodology does not allow one to compare well objects with different $S/N$, nor does it give a statistically meaningful similarity measure. However, it bypasses the difficulties of comparing spectra in data or model space, while producing very robust results.  An additional example for using a distance metric to find similar objects is found in \citet{jofre17}, who looked for nearest neighbors on a t-SNE map of RAVE stars in search for spectroscopic twins. To build the t-SNE map euclidian distance were used.

\section{Efficient outlier detection}
\label{nnws_subsec}
An important usage of the distance matrix is outlier detection. For this purpose we calculate a weirdness score for each object in the sample. This weirdness score is calculated by summing over the distance matrix. In this section we use the t-SNE visualization in order to understand the properties of this weirdness score. We present a new, local, definition of a weirdness score. We use this local weirdness score for APOGEE stars, and find it is more suitable for detecting outliers.

We can use the t-SNE map in order learn about the weirdness score properties. In Figure \ref{subfig:tsne_wall} we color the t-SNE map by the weirdness score. The central, low $\mathrm{W_{all}}$ part of the t-SNE map contains about half of the objects in the sample. These objects are $\mathrm{G}$ and $\mathrm{K}$ giant stars, 
with weak molecular features in their spectra, 
but with prominent metallic features. They comprise one large group of objects with similar spectra. Below we refer to these objects as the main group. Example spectra for such objects are presented in Figure  \ref{fig:specs_most}. For each object in the figure we present the percentile of the object's weirdness score, i.e. the percent of the objects with lower weirdness score.

\begin{figure}
  \includegraphics[width=\columnwidth]{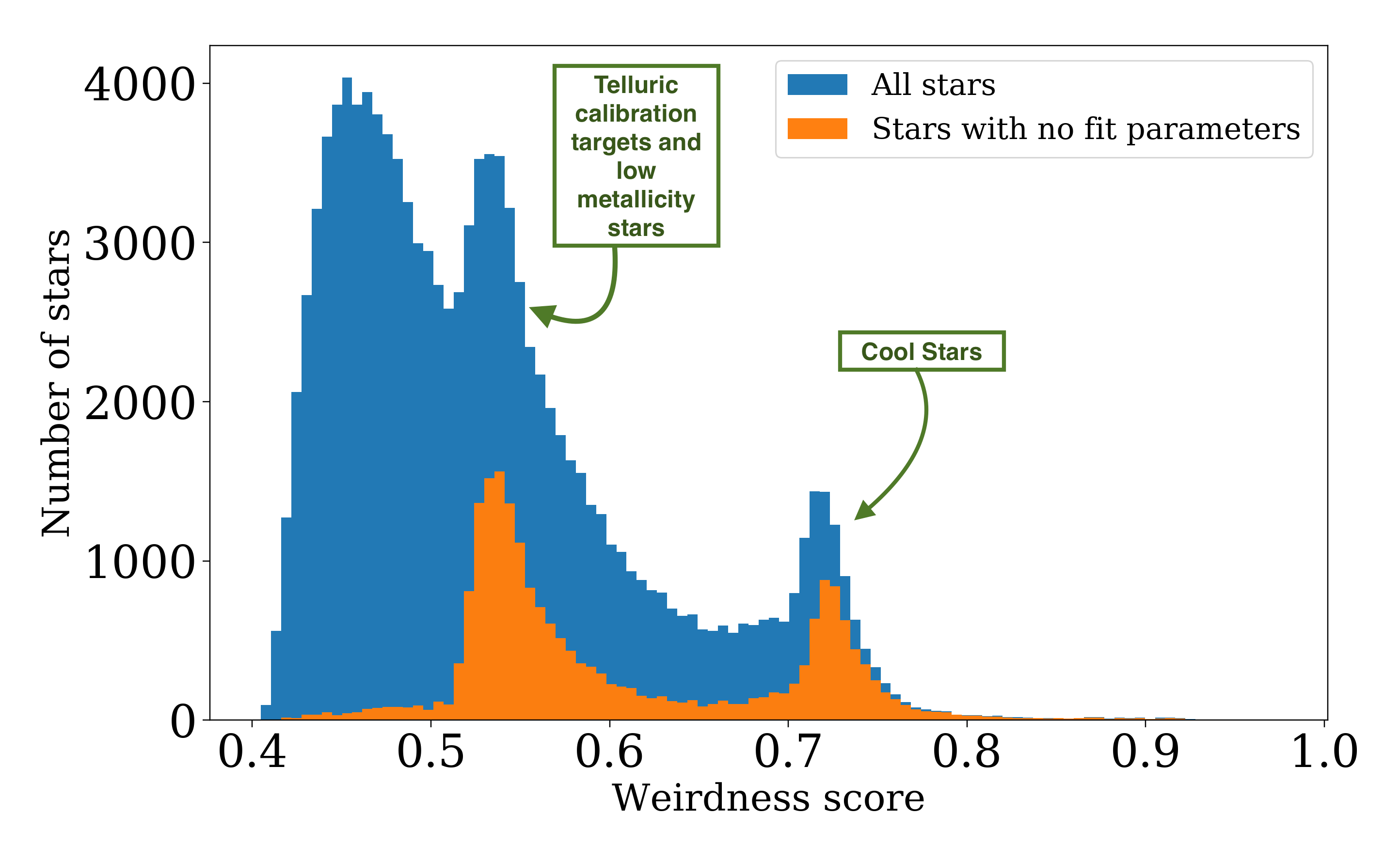}
  \caption{$W_{all}$ distribution for all objects in the sample. The two bumps in the distribution are composed mostly of objects with no ASPCAP fit parameters. Inspecting the spectra of these objects, we see that one bump contains low $\mathrm{T}_{\mathrm{eff}}$ stars, while the other contains low metallicity and hot telluric calibration stars (i.e. stars with weak or non existent absorption lines).}
  \label{fig:wrdns_dist}
 \end{figure}

In order to better understand the properties of $W_{all}$, we examine its distribution in Figure \ref{fig:wrdns_dist}. The distribution decreases smoothly to high weirdness except for two bumps. We interpret the bumps as clusters of stars in our similarity space. One bump consists partially of low-temperature stars, and the other is due to stars with weak or non existent absorption lines - - metal poor stars and telluric calibration targets. 
The bumps in the distribution of $W_{all}$ are due to the fact that there is one dominant cluster of objects in the dataset, and the objects in smaller clusters receive a weirdness score based on how different they are from objects in the main cluster. These results might be useful to detect clusters, or to clean up the dataset from objects outside the main group, but in order to find small classes of interesting outliers, we need a better outlier definition.

To address the issue described above we introduce the 'nearest neighbors weirdness score', a modification to the algorithm that produces 'better outliers' for the APOGEE dataset. When looking for better outliers, we wish to get several different types of objects detected as outliers, in contrast to a weirdness score that strongly correlates to a single attribute (e.g. the effective temperature). In addition we expect to be able to detect known outliers such as binaries and bad spectra. 

\begin{figure}
\begin{center}
  \includegraphics[width=\columnwidth]{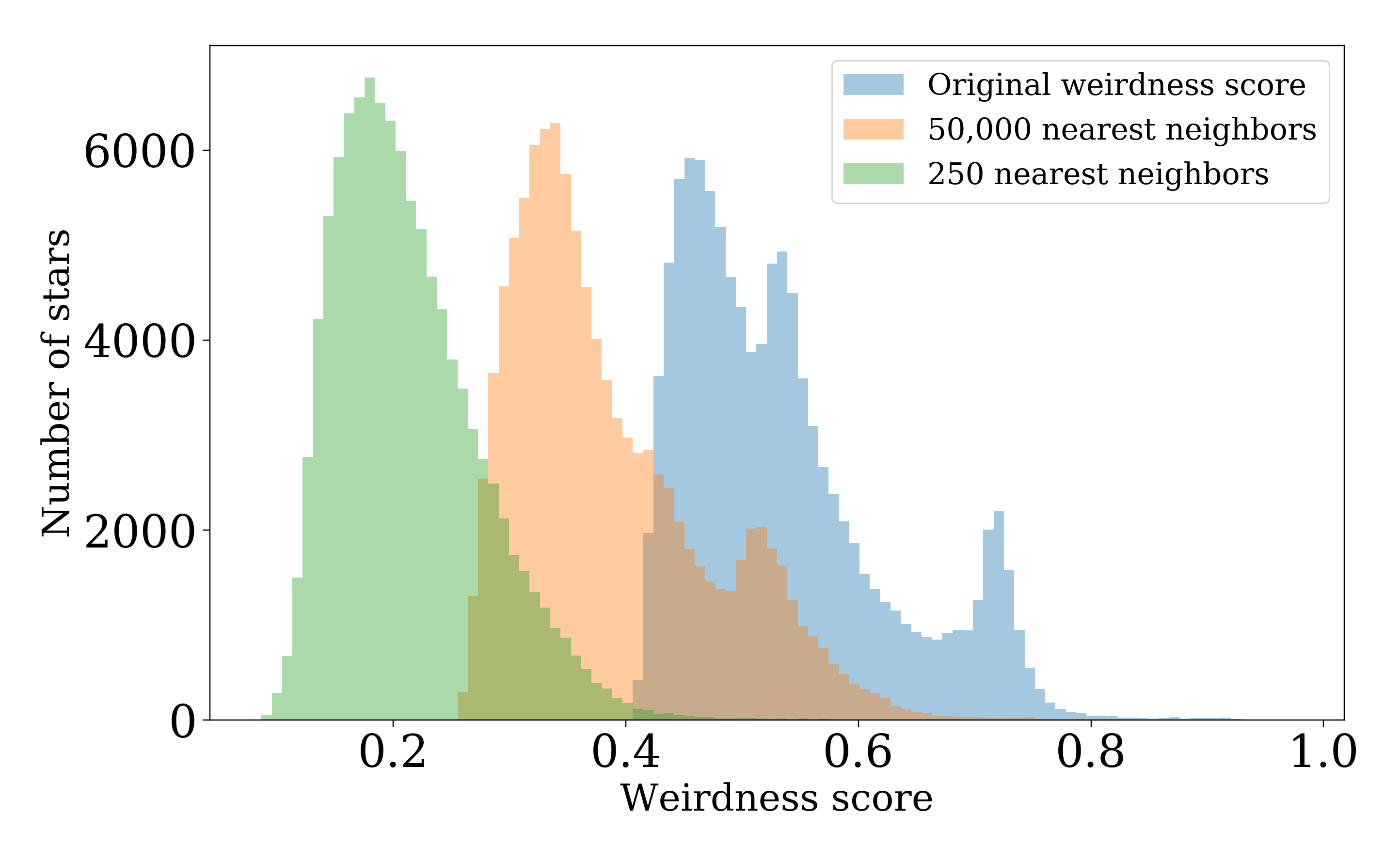}
  \caption{Weirdness score distribution for different numbers of nearest neighbors included in the calculation.}
  \label{fig:wsnn}
  \end{center}
\end{figure}

Instead of defining outliers based on their average distance to the entire sample, we use a more local measure, and for every object we calculate distances to its nearest neighbors. This measure of unusualness is used for distance-based outlier detection in other fields \citep{knorr99,knorr00}. The resulting weirdness score distribution is shown in figure \ref{fig:wsnn}. We can see that the bumps in the weirdness score distribution go away for a small enough number of nearest neighbors. When choosing the number of nearest neighbors to use in the weirdness score calculation, one can check at what point the weirdness score distribution does not contain bumps.

A t-SNE map with the 250 nearest neighbors weirdness score ($W_{250}$) is shown in Figure \ref{subfig:tsne_w250} . Clearly, there is a group of stars that have persistently high (percentile) weirdness score for any number of nearest neighbors used. On the other hand, the high $\mathrm{T}_{\mathrm{eff}}$ stars no longer have high weirdness score for small numbers of nearest neighbors. This results in various other groups of stars receiving higher percentile weirdness score.

An open question regarding many outlier detection algorithms is setting a threshold on the weirdness score, i.e., determining above which weirdness score we mark an object as an "outlier" and inspect it further. The t-SNE map could be of help here too: looking at the $W_{250}$ t-SNE map (Figure \ref{subfig:tsne_w250}) we can see that for each group of stars on the map, the edges receive higher weirdness score. We do not want to mark these edges as outliers, and from the t-SNE map we determine that this would be achieved with a threshold of 0.6, for this specific dataset.

The classification of the outliers is made easier by sorting the objects using their position on the t-SNE map. This way we can classify groups of similar objects instead of one object at a time. Another method we try for outlier inspection is called DEMUD \citep{wagstaff13}. Instead of examining the outliers sequentially, one starts from the weirdest object, and then inspects the weird object that is the farthest from the first, followed by the one farthest from the first two, and so forth. The idea is to sample the different populations of outliers quickly, stopping once we start seeing the same types of objects repeating. For the final classification of the outliers we chose a threshold on the weirdness score (as discussed above) and use the t-SNE map to help with the classification. This is followed by taking a lower threshold on the weirdness score and using DEMUD to look for additional types of outliers. This second step did not result in new types of outliers.

\section{APOGEE Outliers}
\label{resultssec}
\begin{figure}
  \begin{center}
  \includegraphics[width=\columnwidth]{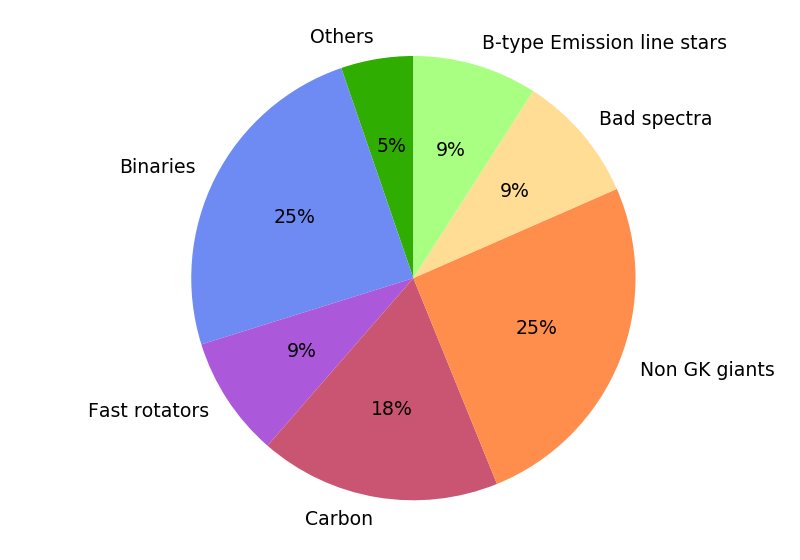}
  \caption{Results of the manual classification of 348 highest $W_{250}$. In the next sections we discuss each of these groups. The Non GK giants group contains mostly M dwarfs. This figure refers to DR13.}
  \label{fig:outclass}
  \end{center}
\end{figure}

In this section we present the results of manual classification of the highest $W_{250}$ stars. Here we use results from both DR14 and DR13, as in DR14 many objects have poorly determined continua. In total we look at 577 objects. The distribution of the different groups of outliers for DR13 only is shown in Figure \ref{fig:outclass}. We find the following large groups: 
Be stars,
young stellar objects,
carbon enriched stars,
 double lined spectroscopic binaries (SB2),
 fast rotators, 
 M dwarfs,
 M giants and cool K giants (these stars have the highest $W_{all}$), 
 and stars with bad spectra.
 In addition we find a number of objects that do not fit into any of the above classes.

``Bad spectra'' are objects with ASPCAP warn flags, combined with a strange looking spectrum. The flags we encounter for the outliers are \textit{commissioning}, \textit{persist high}, and \textit{persist jump neg(high)}. Other "bad spectra" objects are not flagged but have faulty spectra. These appear only in DR14 and we discuss them below.

We note that these classes are not mutually exclusive (in Figure \ref{fig:outclass} each object is assigned to a single class we believe describes it best).

\subsection{B-type emission line stars}
The objects in this group are Be stars. APOGEE targeted approximately 50 known Be stars in an ancillary program, while the additional Be stars in the APOGEE sample were originally targeted as telluric standard stars. 
\citet{chojnowski15,chojnowski17} compiled a catalog of 238 Be stars in the APOGEE dataset. They identified these stars by visual inspection.

We find 40 Be stars not included in the \citet{chojnowski15} catalog. These new Be stars first appeared in DR14. For 26 of these stars emission was never reported before. We list these objects in table \ref{tab:new_be_stars}. Some of these stars were detected as outliers, while the rest were found by inspecting the neighbors, in the distance matrix and t-SNE map, of the outliers.

As seen in Figure \ref{fig:specs_be}, these stars have double peaked H-Br emission lines and weak absorption lines. For some Be stars, metallic emission is also present. ASPCAP fails to derive radial velocities for these objects, due to their unusual spectra.

\subsection{Spectroscopic binaries}
Example spectra of SB2s are shown in Figure \ref{fig:specs_sb2} along with their best fitting synthetic spectra. 
As seen, ASPCAP does not account for binarity.  For some of the SB2s ASPCAP fits broad lines, and for others it fits only one of the two sets of lines. In general the APOGEE reduction pipeline does not have an automatic binary identification routine \citep{nidever15}.

\citet{chojnowski15a} compiled a catalog of spectroscopic binaries in APOGEE \footnote{Their catalog can be found here http://astronomy.nmsu.edu/drewski/apogee-sb2/apSB2.html}. The catalog is constructed by searching for multiple peaks in the spectra cross correlation function, when comparing to the synthetic template spectra. 15 of the 72 binaries we find as outliers are not listed in the catalog of \citet{chojnowski15a} and are therefore new.

\subsection{Fast rotators}
Broad line stars are also detected as outliers. ASPCAP fits broad lines well for dwarf stars but not for giants, as can be seen in Figure \ref{fig:specs_broad}. These stars are all flagged with \textit{suspect broad lines} by the ASPCAP pipeline.

\subsection{Carbon rich stars}
\label{carbon}

Carbon rich stars (discussed in section \ref{sec:tsne_carbon}) are also detected as outliers. In Figure \ref{fig:specs_carbon} we can see the strong $\mathrm{CN}$ compared to $\mathrm{OH}$ lines for a few carbon enriched stars. The weirdness score increases with the strength of the $\mathrm{CN}$ features. 

\subsection{Young stellar objects}
Stars in this group show both H-Br emission lines as well as regular metallic absorption lines. They are mostly young stars included in the INfrared Survey of Young Nebulous Clusters \citep[IN-SYNC, ][]{cottaar14}.
We detect stars with both broad and narrow emission, and with absorption that can be broad or narrow as well as double lined (SB2s).  SIMBAD classification for stars in this group include  'Variable star of Orion type',  'T Tau-type Star', 'Pre-main sequence Star', and  'Young stellar object'.

\subsection{M dwarfs}
M dwarf stars are also detected as outliers. This is due to the small number of M dwarf stars in the APOGEE sample. Example spectra are in Figure \ref{fig:specs_mdwarf}.

\subsection{Other outliers}
Some of the objects detected as outliers did not fall into any of the above classes. These include a brown dwarf, a Wolf-Rayet star, a few AGB stars including an OH-IR star, and known variable stars, as well as three red supergiants observed in the massive stars ancillary program. Also detected as outliers are special non stellar targets, such as the center of M32, a few M31 globular clusters, and three planetary nebulae.

Two outliers show double peaked H-Br emission lines, as well as absorption lines typical to the APOGEE dataset. Both of these objects show RV modulations, suggesting they are multiple star systems. For the first, 2M04052624+5304494, the RV modulation of the absorption lines (determined from APOGEE visit spectra), could be modeled with a period of $\mathrm{P} = 11.152 \pm 0.072 \; \mathrm{days}$, and amplitude of $\mathrm{K} = 78 \pm 15 \;  \mathrm{km} \; \mathrm{s}^{-1}$. The emission lines in the APOGEE spectra show smaller RV modulation, if any.
For the second, 2M06415063-0130177, the absorption lines RV changes by $\sim 160 \; \mathrm{km} \; \mathrm{s}^{-1}$ between two APOGEE visits. The visits are separated by 28 days. For the emission lines we could not get a good estimate on the RVs, as the emission line profiles change significantly between the visits. A CoRoT light-curve is available for this system, showing clear periodic modulation. A period of $\mathrm{P} = 29.04 \; \mathrm{days}$ was derived for this light-curve by \citet{affer12}. We note that this period does not agree with the RV modulation. For both of these systems additional work is required to determine their nature. 

We also detect a group of objects with similar, very broad features. Most of these objects have SIMBAD classifications as contact binaries, mainly W Ursae Majoris.

\begin{figure*}
\includegraphics[width=1.7\columnwidth]{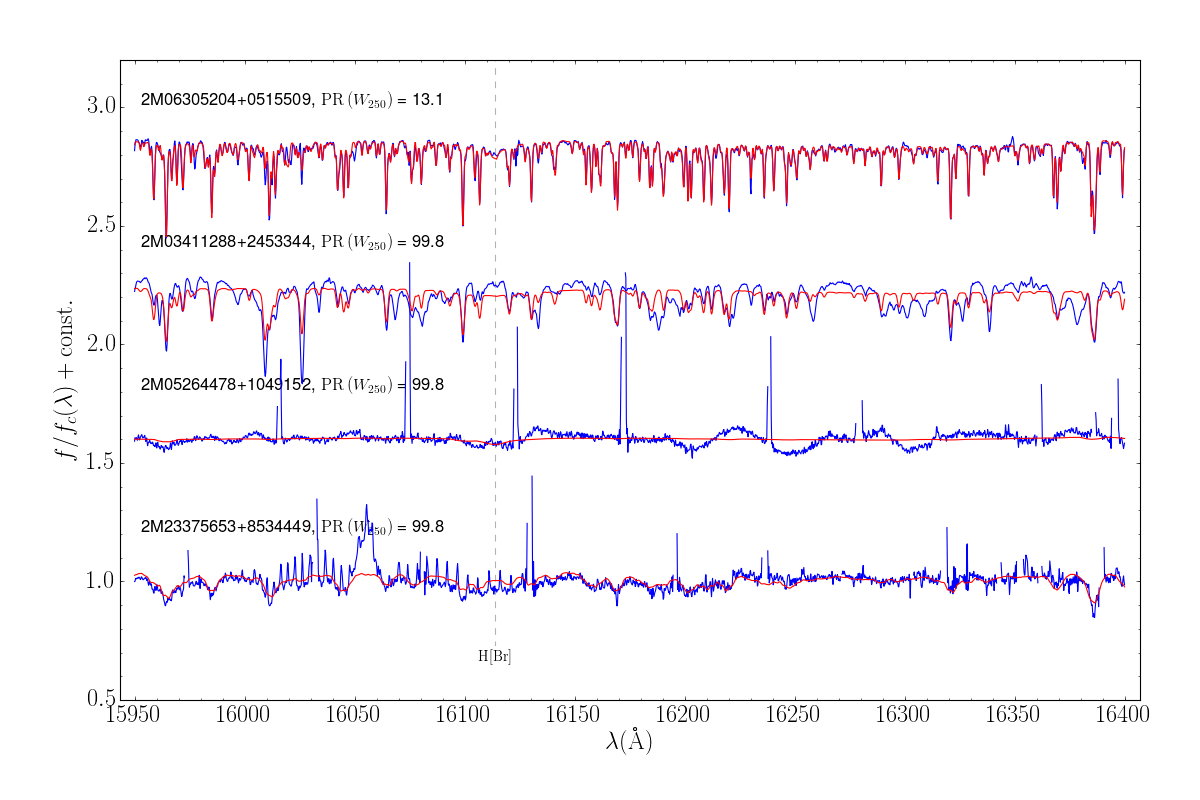}
\caption{Spectra for the three unclassified outliers. Top spectrum is a typical APOGEE red giant, for comparison. The red line is the ASPCAP fit and the blue line is the PCN spectrum, except where indicated. $\mathrm{PR}\left(W_{250}\right)$ indicates the percentile of objects with lower weirdness score. Relative fluxes are offset by a constant for display purposes.}
\label{fig:unclassified}
\end{figure*}

A few objects remain unexplained. We divide these objects into two groups. In the first group we have objects with spectra that seems to have similar features to typical APOGEE red giants (by means of visual inspection). The second group contains stars with spectra that are clearly different from typical APOGEE red giants. We refer to the first group as unexplained red giants, and to the second group as unexplained non red giants. Some of the unexplained red giants stars have low carbon and high nitrogen ASPCAP abundances. Inspecting their spectra, we do see significantly weaker CO features relative to low weirdness score stars with similar stellar parameters. One of these stars, 2M17534571-2949362, is discussed in \cite{fernandez-trincado17} as having low Mg, but high Al and N abundances. There are three unexplained non red giants, the first is 2M03411288+2453344, which was targeted as a telluric calibrator target. As could be seen in Figure \ref{fig:specs_other}, the ASPCAP fit does not catch many of the features in the spectrum, in particular there is no H-Br absorption. The cross correlation function shows a single peak, suggesting it is not a binary star. There are 3 visits to this star, all showing the same features. The objects most similar to this one, according to the distance matrix, do not show similar features. 2M05264478+1049152 has very broad features that we do not identify. Same goes for 2M23375653+8534449 which also has a single emission line centered at $\lambda = 16055 [\angstrom]$ that we cannot identify. We show the spectra of the unexplained non red giants in Figure \ref{fig:unclassified}.

\subsection{Bad reductions}
In DR14 roughly half of the high weirdness score objects have badly determined continua. We show a few examples in Figure \ref{fig:specs}. These objects can be divided into two groups. For the first group, the issue seems to be a bug in the ASPCAP PCN process. For objects in this group the combined unnormalized spectra looks regular, as well as the DR13 PCN spectra (for objects with available DR13 data). For the second group already at least one of the visit spectra is faulty, and this error propagates down the pipeline. Examples for both of these errors are presented in Figure \ref{fig:specs_bad}.

\section{Summary}
\label{sec:sum}
In this work we calculate a similarity measure for APOGEE infrared spectra of stars. We show that this similarity matrix traces physical properties such as effective temperature, metallicity and surface gravity.  Such a similarity matrix could be used for object retrieval, i.e., finding objects that are similar to a given example, it can be used to detect outliers, and more generally to assist learning about the structure of a dataset. The similarity is obtained without inputing information derived by model fitting, and thus the similarity could be used to query and learn about objects that are not well fitted by the pipeline and as such are hard to find using the fit parameters database.

As noted above, we find that the unsupervised RF is capable of aggregating complex spectral information into a single number, the pair-wise distance between two objects. We find that various stellar parameters are encoded into this distance, and that the resulting RF represents a general model of stellar spectra (\cite{baron17a} showed that this is true for spectra of galaxies). As such, one can imagine inverting the process, and using the trained RF to generate ``real-looking" objects, which is in turn a generative model.

Using this unsupervised RF distance matrix and dimensionality reduction techniques, one can study the structure of the data, and the relations between different classes of objects within a dataset. However, it is worth noting that much of the insight gained about the APOGEE sample was made possible using the derived ASPCAP stellar parameters. Without these labels, coloring the t-SNE map would not have been possible. While our proposed unsupervised distance matrix contains various types of information, the extraction of this information still heavily depends on annotations of the distance matrix. Thus, for many applications, it is only the combination of our approach and existing knowledge about the dataset that can be useful to gain additional insight. 

Using our distance matrix to detect outliers, we find objects from the following types of known classes: B-type emission-line stars, carbon rich stars, spectroscopic binaries, broad line stars, young stars, bad spectra, and M dwarfs (which are ordinary but underrepresented in the dataset), showing that the algorithm is capable of detecting a wide variety of phenomena. A few dozens of objects that were detected as outliers did not fall into any of the large groups, these include special targets such as galaxies, globular clusters, and planetary nebulae, stars with unusual abundances, contact binaries, stars observed with the massive star ancillary program and more. Three outliers remain without explanation.

Some of the carbon rich outliers have a poor ASPCAP fit, though these groups are included in the ASPCAP stellar spectral library. Possibly the objects without a good fit are extreme cases and could be used to improve and test the pipeline. 
The SB2s detected as outliers have diverse types of spectra and could be used to test SB2 detection specific algorithms. Bad spectra objects and underrepresented objects are not interesting by themselves, but detecting them could be useful in order to clean the sample and find bugs in the pipeline. Finding new Be stars is an example for detection of new objects of known types using the distance matrix or t-SNE map. This is especially useful in larger surveys, where visual inspection is not feasible.

The use of t-SNE to visualize the distance matrix was also useful for the purpose of outlier detection.  This enabled us to speed up the classification of the outliers by classifying nearby objects together. More importantly, the t-SNE map proved to be useful in learning about the regular objects in the data set, an important step to take before looking at the outliers. Viewing spectra of objects located in different regions of the t-SNE map allowed us to quickly review the different classes of regular objects. For the APOGEE dataset, in which there is one large group of similar objects, a nearest neighbors weirdness score, or a 'local' weirdness score, was needed in order to detect the interesting outliers. Although this was not required to detect the interesting outlying galaxies in \citet{baron17a}, we believe the the local weirdness score is more general and should be used in future work. The number of nearest neighbors to use when calculating the local weirdness score is dataset dependent.  Coloring the t-SNE map by the different weirdness scores or building t-SNE maps with different perplexities, can help decide on which number of nearest neighbors is appropriate. It is also possible that in order to detect all interesting objects one type of nearest neighbors weirdness score would not be enough, as different types of outliers can come in different (small) cluster sizes. In our case the outliers population seemed robust to a number of nearest neighbors from a few to a few thousands. We note that for the map shown in Figure \ref{fig:t_tsne} we used perplexity of 2000. This value was chosen in order to make the visualization relatively simple. With smaller perplexity we obtained maps with more complex small scale structure, such as small clusters. These maps could be useful for investigating the data further but for a clean visualization of the large scale structure we used a high perplexity map.

Future work could involve combining the distance matrix, which is based on spectral data alone, with other types of available data. A natural direction is the physical position of a star. For example, one can look for stars that are normal compared to the entire population of stars, but are weird when compared to their local environment. A table with the 100 nearest neighbors of each object, including their respective distances, is available online. We also include the coordinates for the t-SNE map shown above. Examples for using these data products are available in a Jupyter Notebook \footnote{\href{https://github.com/ireis/APOGEE\_tSNE\_nb}{github.com/ireis/APOGEE\_tSNE\_nb}}.

\section*{Acknowledgements}
We thank D. Hogg for suggesting the use of t-SNE, and other useful comments, and D. Chojnowski for discussing some of the outliers. We also thank the reviewer for helpful suggestions to improve this manuscript.

This research made use of: the NASA Astrophysics Data System Bibliographic Services, scikit-learn \citep[][]{pedregosa11}, SciPy \citep[][]{scipy01},  IPython \citep[][]{perez07}, matplotlib \citep[][]{hunter07}, astropy \citep[][]{astropy-collaboration13} and the SIMBAD database \citep[][]{wenger00}.

This work made extensive use of SDSS data. Funding for the Sloan Digital Sky Survey IV has been provided by the Alfred P. Sloan Foundation, the U.S. Department of Energy Office of Science, and the Participating Institutions. SDSS-IV acknowledges
support and resources from the Center for High-Performance Computing at
the University of Utah. The SDSS web site is www.sdss.org.

SDSS-IV is managed by the Astrophysical Research Consortium for the 
Participating Institutions of the SDSS Collaboration including the 
Brazilian Participation Group, the Carnegie Institution for Science, 
Carnegie Mellon University, the Chilean Participation Group, the French Participation Group, Harvard-Smithsonian Center for Astrophysics, 
Instituto de Astrof\'isica de Canarias, The Johns Hopkins University, 
Kavli Institute for the Physics and Mathematics of the Universe (IPMU) / 
University of Tokyo, Lawrence Berkeley National Laboratory, 
Leibniz Institut f\"ur Astrophysik Potsdam (AIP),  
Max-Planck-Institut f\"ur Astronomie (MPIA Heidelberg), 
Max-Planck-Institut f\"ur Astrophysik (MPA Garching), 
Max-Planck-Institut f\"ur Extraterrestrische Physik (MPE), 
National Astronomical Observatories of China, New Mexico State University, 
New York University, University of Notre Dame, 
Observat\'ario Nacional / MCTI, The Ohio State University, 
Pennsylvania State University, Shanghai Astronomical Observatory, 
United Kingdom Participation Group,
Universidad Nacional Aut\'onoma de M\'exico, University of Arizona, 
University of Colorado Boulder, University of Oxford, University of Portsmouth, 
University of Utah, University of Virginia, University of Washington, University of Wisconsin, 
Vanderbilt University, and Yale University.




\bibliographystyle{mnras}
\bibliography{apogee_2} 

\begin{thebibliography}{}
\makeatletter
\relax
\def\mn@urlcharsother{\let\do\@makeother \do\$\do\&\do\#\do\^\do\_\do\%\do\~}
\def\mn@doi{\begingroup\mn@urlcharsother \@ifnextchar [ {\mn@doi@}
  {\mn@doi@[]}}
\def\mn@doi@[#1]#2{\def\@tempa{#1}\ifx\@tempa\@empty \href
  {http://dx.doi.org/#2} {doi:#2}\else \href {http://dx.doi.org/#2} {#1}\fi
  \endgroup}
\def\mn@eprint#1#2{\mn@eprint@#1:#2::\@nil}
\def\mn@eprint@arXiv#1{\href {http://arxiv.org/abs/#1} {{\tt arXiv:#1}}}
\def\mn@eprint@dblp#1{\href {http://dblp.uni-trier.de/rec/bibtex/#1.xml}
  {dblp:#1}}
\def\mn@eprint@#1:#2:#3:#4\@nil{\def\@tempa {#1}\def\@tempb {#2}\def\@tempc
  {#3}\ifx \@tempc \@empty \let \@tempc \@tempb \let \@tempb \@tempa \fi \ifx
  \@tempb \@empty \def\@tempb {arXiv}\fi \@ifundefined
  {mn@eprint@\@tempb}{\@tempb:\@tempc}{\expandafter \expandafter \csname
  mn@eprint@\@tempb\endcsname \expandafter{\@tempc}}}

\bibitem[\protect\citeauthoryear{{Abolfathi} et~al.,}{{Abolfathi}
  et~al.}{2017}]{abolfathi17}
{Abolfathi} B.,  et~al., 2017, preprint, \href
  {http://adsabs.harvard.edu/abs/2017arXiv170709322A} {} (\mn@eprint {arXiv}
  {1707.09322})

\bibitem[\protect\citeauthoryear{{Affer}, {Micela}, {Favata}  \&
  {Flaccomio}}{{Affer} et~al.}{2012}]{affer12}
{Affer} L.,  {Micela} G.,  {Favata} F.,   {Flaccomio} E.,  2012, \mn@doi
  [\mnras] {10.1111/j.1365-2966.2012.20802.x}, \href
  {http://adsabs.harvard.edu/abs/2012MNRAS.424...11A} {424, 11}

\bibitem[\protect\citeauthoryear{{Astropy Collaboration} et~al.,}{{Astropy
  Collaboration} et~al.}{2013}]{astropy-collaboration13}
{Astropy Collaboration} et~al., 2013, \mn@doi [\aap]
  {10.1051/0004-6361/201322068}, \href
  {http://adsabs.harvard.edu/abs/2013A%26A...558A..33A} {558, A33}

\bibitem[\protect\citeauthoryear{{Ball} \& {Brunner}}{{Ball} \&
  {Brunner}}{2010}]{ball10}
{Ball} N.~M.,  {Brunner} R.~J.,  2010, \mn@doi [International Journal of Modern
  Physics D] {10.1142/S0218271810017160}, \href
  {http://adsabs.harvard.edu/abs/2010IJMPD..19.1049B} {19, 1049}

\bibitem[\protect\citeauthoryear{{Baron} \& {Poznanski}}{{Baron} \&
  {Poznanski}}{2017}]{baron17a}
{Baron} D.,  {Poznanski} D.,  2017, \mn@doi [\mnras] {10.1093/mnras/stw3021},
  \href {http://adsabs.harvard.edu/abs/2017MNRAS.465.4530B} {465, 4530}

\bibitem[\protect\citeauthoryear{{Baron}, {Poznanski}, {Watson}, {Yao}, {Cox}
  \& {Prochaska}}{{Baron} et~al.}{2015}]{baron15}
{Baron} D.,  {Poznanski} D.,  {Watson} D.,  {Yao} Y.,  {Cox} N.~L.~J.,
  {Prochaska} J.~X.,  2015, \mn@doi [\mnras] {10.1093/mnras/stv977}, \href
  {http://adsabs.harvard.edu/abs/2015MNRAS.451..332B} {451, 332}

\bibitem[\protect\citeauthoryear{{Baron}, {Netzer}, {Poznanski}, {Prochaska}
  \& {F{\"o}rster Schreiber}}{{Baron} et~al.}{2017}]{baron17}
{Baron} D.,  {Netzer} H.,  {Poznanski} D.,  {Prochaska} J.~X.,   {F{\"o}rster
  Schreiber} N.~M.,  2017, \mn@doi [\mnras] {10.1093/mnras/stx1329}, \href
  {http://adsabs.harvard.edu/abs/2017MNRAS.470.1687B} {470, 1687}

\bibitem[\protect\citeauthoryear{{Bloom} et~al.,}{{Bloom}
  et~al.}{2012}]{bloom12}
{Bloom} J.~S.,  et~al., 2012, \mn@doi [\pasp] {10.1086/668468}, \href
  {http://adsabs.harvard.edu/abs/2012PASP..124.1175B} {124, 1175}

\bibitem[\protect\citeauthoryear{{Bovy}}{{Bovy}}{2016}]{bovy16}
{Bovy} J.,  2016, \mn@doi [\apj] {10.3847/0004-637X/817/1/49}, \href
  {http://adsabs.harvard.edu/abs/2016ApJ...817...49B} {817, 49}

\bibitem[\protect\citeauthoryear{{Bovy} et~al.,}{{Bovy} et~al.}{2014}]{bovy14}
{Bovy} J.,  et~al., 2014, \mn@doi [\apj] {10.1088/0004-637X/790/2/127}, \href
  {http://adsabs.harvard.edu/abs/2014ApJ...790..127B} {790, 127}

\bibitem[\protect\citeauthoryear{Breiman}{Breiman}{2001}]{breiman01}
Breiman L.,  2001, \mn@doi [Machine Learning] {10.1023/A:1010933404324}, 45, 5

\bibitem[\protect\citeauthoryear{Breiman \& Cutler}{Breiman \&
  Cutler}{2003}]{breiman03}
Breiman L.,  Cutler A.,  2003, Thechnical Report

\bibitem[\protect\citeauthoryear{Breiman, Friedman, Olshen  \& Stone}{Breiman
  et~al.}{1984}]{breiman84}
Breiman L.,  Friedman J.~H.,  Olshen R.~A.,   Stone C.~J.,  1984, -

\bibitem[\protect\citeauthoryear{{Chiappini} et~al.,}{{Chiappini}
  et~al.}{2015}]{chiappini15}
{Chiappini} C.,  et~al., 2015, \mn@doi [\aap] {10.1051/0004-6361/201525865},
  \href {http://adsabs.harvard.edu/abs/2015A%26A...576L..12C} {576, L12}

\bibitem[\protect\citeauthoryear{{Chojnowski} et~al.,}{{Chojnowski}
  et~al.}{2015a}]{chojnowski15}
{Chojnowski} S.~D.,  et~al., 2015a, \mn@doi [\aj] {10.1088/0004-6256/149/1/7},
  \href {http://adsabs.harvard.edu/abs/2015AJ....149....7C} {149, 7}

\bibitem[\protect\citeauthoryear{{Chojnowski} et~al.,}{{Chojnowski}
  et~al.}{2015b}]{chojnowski15a}
{Chojnowski} S.~D.,  et~al., 2015b, in American Astronomical Society Meeting
  Abstracts. p. 340.05

\bibitem[\protect\citeauthoryear{{Chojnowski} et~al.,}{{Chojnowski}
  et~al.}{2017}]{chojnowski17}
{Chojnowski} S.~D.,  et~al., 2017, \mn@doi [\aj] {10.3847/1538-3881/aa64ce},
  \href {http://adsabs.harvard.edu/abs/2017AJ....153..174C} {153, 174}

\bibitem[\protect\citeauthoryear{{Cottaar} et~al.,}{{Cottaar}
  et~al.}{2014}]{cottaar14}
{Cottaar} M.,  et~al., 2014, \mn@doi [\apj] {10.1088/0004-637X/794/2/125},
  \href {http://adsabs.harvard.edu/abs/2014ApJ...794..125C} {794, 125}

\bibitem[\protect\citeauthoryear{{Eisenstein} et~al.,}{{Eisenstein}
  et~al.}{2011}]{eisenstein11}
{Eisenstein} D.~J.,  et~al., 2011, \mn@doi [\aj] {10.1088/0004-6256/142/3/72},
  \href {http://adsabs.harvard.edu/abs/2011AJ....142...72E} {142, 72}

\bibitem[\protect\citeauthoryear{{Fern{\'a}ndez-Trincado}
  et~al.,}{{Fern{\'a}ndez-Trincado} et~al.}{2017}]{fernandez-trincado17}
{Fern{\'a}ndez-Trincado} J.~G.,  et~al., 2017, \mn@doi [\apjl]
  {10.3847/2041-8213/aa8032}, \href
  {http://adsabs.harvard.edu/abs/2017ApJ...846L...2F} {846, L2}

\bibitem[\protect\citeauthoryear{{Frinchaboy} et~al.,}{{Frinchaboy}
  et~al.}{2013}]{frinchaboy13}
{Frinchaboy} P.~M.,  et~al., 2013, \mn@doi [\apjl]
  {10.1088/2041-8205/777/1/L1}, \href
  {http://adsabs.harvard.edu/abs/2013ApJ...777L...1F} {777, L1}

\bibitem[\protect\citeauthoryear{{Garcia-Dias}, {Allende Prieto}, {S{\'a}nchez
  Almeida}  \& {Ordov{\'a}s-Pascual}}{{Garcia-Dias}
  et~al.}{2018}]{garcia-dias18}
{Garcia-Dias} R.,  {Allende Prieto} C.,  {S{\'a}nchez Almeida} J.,
  {Ordov{\'a}s-Pascual} I.,  2018, preprint, \href
  {http://adsabs.harvard.edu/abs/2018arXiv180107912G} {} (\mn@eprint {arXiv}
  {1801.07912})

\bibitem[\protect\citeauthoryear{{Garc{\'{\i}}a P{\'e}rez}
  et~al.,}{{Garc{\'{\i}}a P{\'e}rez} et~al.}{2016}]{garcia-perez16}
{Garc{\'{\i}}a P{\'e}rez} A.~E.,  et~al., 2016, \mn@doi [\aj]
  {10.3847/0004-6256/151/6/144}, \href
  {http://adsabs.harvard.edu/abs/2016AJ....151..144G} {151, 144}

\bibitem[\protect\citeauthoryear{{Hayden} et~al.,}{{Hayden}
  et~al.}{2015}]{hayden15}
{Hayden} M.~R.,  et~al., 2015, \mn@doi [\apj] {10.1088/0004-637X/808/2/132},
  \href {http://adsabs.harvard.edu/abs/2015ApJ...808..132H} {808, 132}

\bibitem[\protect\citeauthoryear{Hunter}{Hunter}{2007}]{hunter07}
Hunter J.~D.,  2007, \mn@doi [Computing In Science \& Engineering]
  {10.1109/MCSE.2007.55}, 9, 90

\bibitem[\protect\citeauthoryear{{Jofr{\'e}}, {M{\"a}dler}, {Gilmore}, {Casey},
  {Soubiran}  \& {Worley}}{{Jofr{\'e}} et~al.}{2015}]{jofre15}
{Jofr{\'e}} P.,  {M{\"a}dler} T.,  {Gilmore} G.,  {Casey} A.~R.,  {Soubiran}
  C.,   {Worley} C.,  2015, \mn@doi [\mnras] {10.1093/mnras/stv1724}, \href
  {http://adsabs.harvard.edu/abs/2015MNRAS.453.1428J} {453, 1428}

\bibitem[\protect\citeauthoryear{{Jofr{\'e}} et~al.,}{{Jofr{\'e}}
  et~al.}{2017}]{jofre17}
{Jofr{\'e}} P.,  et~al., 2017, \mn@doi [\mnras] {10.1093/mnras/stx1877}, \href
  {http://adsabs.harvard.edu/abs/2017MNRAS.472.2517J} {472, 2517}

\bibitem[\protect\citeauthoryear{Jones, Oliphant, Peterson  et~al.}{Jones
  et~al.}{01  }]{scipy01}
Jones E.,  Oliphant T.,  Peterson P.,   et~al., 2001--, {SciPy}: Open source
  scientific tools for {Python}, \url {http://www.scipy.org/}

\bibitem[\protect\citeauthoryear{Knorr \& Ng}{Knorr \& Ng}{1999}]{knorr99}
Knorr E.~M.,  Ng R.~T.,  1999, in Proceedings of the 25th International
  Conference on Very Large Data Bases. VLDB '99.
Morgan Kaufmann Publishers Inc., San Francisco, CA, USA, pp 211--222, \url
  {http://dl.acm.org/citation.cfm?id=645925.671529}

\bibitem[\protect\citeauthoryear{Knorr, Ng  \& Tucakov}{Knorr
  et~al.}{2000}]{knorr00}
Knorr E.~M.,  Ng R.~T.,   Tucakov V.,  2000, \mn@doi [The VLDB Journal]
  {10.1007/s007780050006}, 8, 237

\bibitem[\protect\citeauthoryear{{Majewski}, {APOGEE Team}  \& {APOGEE-2
  Team}}{{Majewski} et~al.}{2016}]{majewski16}
{Majewski} S.~R.,  {APOGEE Team}  {APOGEE-2 Team} 2016, \mn@doi [Astronomische
  Nachrichten] {10.1002/asna.201612387}, \href
  {http://adsabs.harvard.edu/abs/2016AN....337..863M} {337, 863}

\bibitem[\protect\citeauthoryear{{Masci}, {Hoffman}, {Grillmair}  \&
  {Cutri}}{{Masci} et~al.}{2014}]{masci14}
{Masci} F.~J.,  {Hoffman} D.~I.,  {Grillmair} C.~J.,   {Cutri} R.~M.,  2014,
  \mn@doi [\aj] {10.1088/0004-6256/148/1/21}, \href
  {http://adsabs.harvard.edu/abs/2014AJ....148...21M} {148, 21}

\bibitem[\protect\citeauthoryear{{Meusinger}, {Schalldach}, {Scholz}, {in der
  Au}, {Newholm}, {de Hoon}  \& {Kaminsky}}{{Meusinger}
  et~al.}{2012}]{meusinger12}
{Meusinger} H.,  {Schalldach} P.,  {Scholz} R.-D.,  {in der Au} A.,  {Newholm}
  M.,  {de Hoon} A.,   {Kaminsky} B.,  2012, \mn@doi [\aap]
  {10.1051/0004-6361/201118143}, \href
  {http://adsabs.harvard.edu/abs/2012A%26A...541A..77M} {541, A77}

\bibitem[\protect\citeauthoryear{{Miller}, {Kulkarni}, {Cao}, {Laher}, {Masci}
  \& {Surace}}{{Miller} et~al.}{2017}]{miller17}
{Miller} A.~A.,  {Kulkarni} M.~K.,  {Cao} Y.,  {Laher} R.~R.,  {Masci} F.~J.,
  {Surace} J.~A.,  2017, \mn@doi [\aj] {10.3847/1538-3881/153/2/73}, \href
  {http://adsabs.harvard.edu/abs/2017AJ....153...73M} {153, 73}

\bibitem[\protect\citeauthoryear{{Ness}, {Hogg}, {Rix}, {Ho}  \&
  {Zasowski}}{{Ness} et~al.}{2015}]{ness15}
{Ness} M.,  {Hogg} D.~W.,  {Rix} H.-W.,  {Ho} A.~Y.~Q.,   {Zasowski} G.,  2015,
  \mn@doi [\apj] {10.1088/0004-637X/808/1/16}, \href
  {http://adsabs.harvard.edu/abs/2015ApJ...808...16N} {808, 16}

\bibitem[\protect\citeauthoryear{{Nidever} et~al.,}{{Nidever}
  et~al.}{2014}]{nidever14}
{Nidever} D.~L.,  et~al., 2014, \mn@doi [\apj] {10.1088/0004-637X/796/1/38},
  \href {http://adsabs.harvard.edu/abs/2014ApJ...796...38N} {796, 38}

\bibitem[\protect\citeauthoryear{{Nidever} et~al.,}{{Nidever}
  et~al.}{2015}]{nidever15}
{Nidever} D.~L.,  et~al., 2015, \mn@doi [\aj] {10.1088/0004-6256/150/6/173},
  \href {http://adsabs.harvard.edu/abs/2015AJ....150..173N} {150, 173}

\bibitem[\protect\citeauthoryear{Pedregosa et~al.,}{Pedregosa
  et~al.}{2011}]{pedregosa11}
Pedregosa F.,  et~al., 2011, Journal of Machine Learning Research, 12, 2825

\bibitem[\protect\citeauthoryear{P\'erez \& Granger}{P\'erez \&
  Granger}{2007}]{perez07}
P\'erez F.,  Granger B.~E.,  2007, \mn@doi [Computing in Science and
  Engineering] {10.1109/MCSE.2007.53}, 9, 21

\bibitem[\protect\citeauthoryear{Pimentel, Clifton, Clifton  \&
  Tarassenko}{Pimentel et~al.}{2014}]{pimentel14}
Pimentel M.~A.,  Clifton D.~A.,  Clifton L.,   Tarassenko L.,  2014, \mn@doi
  [Signal Processing] {https://doi.org/10.1016/j.sigpro.2013.12.026}, 99, 215

\bibitem[\protect\citeauthoryear{{Schawinski}, {Zhang}, {Zhang}, {Fowler}  \&
  {Santhanam}}{{Schawinski} et~al.}{2017}]{schawinski17}
{Schawinski} K.,  {Zhang} C.,  {Zhang} H.,  {Fowler} L.,   {Santhanam} G.~K.,
  2017, \mn@doi [\mnras] {10.1093/mnrasl/slx008}, \href
  {http://adsabs.harvard.edu/abs/2017MNRAS.467L.110S} {467, L110}

\bibitem[\protect\citeauthoryear{Shi \& Horvath}{Shi \& Horvath}{2006}]{shi06}
Shi T.,  Horvath S.,  2006, \mn@doi [Journal of Computational and Graphical
  Statistics] {10.1198/106186006X94072}, 15, 118

\bibitem[\protect\citeauthoryear{Wagstaff, Lanza, Thompson, Dietterich  \&
  Gilmore}{Wagstaff et~al.}{2013}]{wagstaff13}
Wagstaff K.~L.,  Lanza N.~L.,  Thompson D.~R.,  Dietterich T.~G.,   Gilmore
  M.~S.,  2013, in Proceedings of the Twenty-Seventh AAAI Conference on
  Artificial Intelligence. AAAI'13.
AAAI Press, pp 905--911, \url
  {http://dl.acm.org/citation.cfm?id=2891460.2891586}

\bibitem[\protect\citeauthoryear{Wattenberg, Vi{\'e}gas  \& Johnson}{Wattenberg
  et~al.}{2016}]{wattenberg2016how}
Wattenberg M.,  Vi{\'e}gas F.,   Johnson I.,  2016, \mn@doi [Distill]
  {10.23915/distill.00002}

\bibitem[\protect\citeauthoryear{{Wenger} et~al.,}{{Wenger}
  et~al.}{2000}]{wenger00}
{Wenger} M.,  et~al., 2000, \mn@doi [\aaps] {10.1051/aas:2000332}, \href
  {http://adsabs.harvard.edu/abs/2000A%26AS..143....9W} {143, 9}

\bibitem[\protect\citeauthoryear{Yang}{Yang}{2006}]{yang06}
Yang L.,  2006, Distance Metric Learning: A Comprehensive Survey

\bibitem[\protect\citeauthoryear{{Zasowski} et~al.,}{{Zasowski}
  et~al.}{2013}]{zasowski13}
{Zasowski} G.,  et~al., 2013, \mn@doi [\aj] {10.1088/0004-6256/146/4/81}, \href
  {http://adsabs.harvard.edu/abs/2013AJ....146...81Z} {146, 81}

\bibitem[\protect\citeauthoryear{{Zasowski} et~al.,}{{Zasowski}
  et~al.}{2017}]{zasowski17}
{Zasowski} G.,  et~al., 2017, preprint, \href
  {http://adsabs.harvard.edu/abs/2017arXiv170800155Z} {} (\mn@eprint {arXiv}
  {1708.00155})

\bibitem[\protect\citeauthoryear{van~der Maaten \& Hinton}{van~der Maaten \&
  Hinton}{2008}]{maaten08}
van~der Maaten L.,  Hinton G.,  2008, -

\makeatother
\end{thebibliography}




\appendix

\section{Spectra and tables}

In Figure \ref{fig:specs} we show example spectra of objects from the different outlying groups, as well as spectroscopic twins.
\begin{figure*}
\centering
	\subfloat[Regular objects, i.e. objects with low weirdness scores. Clearly, all have similar spectra. $\mathrm{PR}\left(W_{250}\right)$ indicates the percentile of objects with lower weirdness score.]{%
    \includegraphics[width=\specfigwidth\columnwidth]{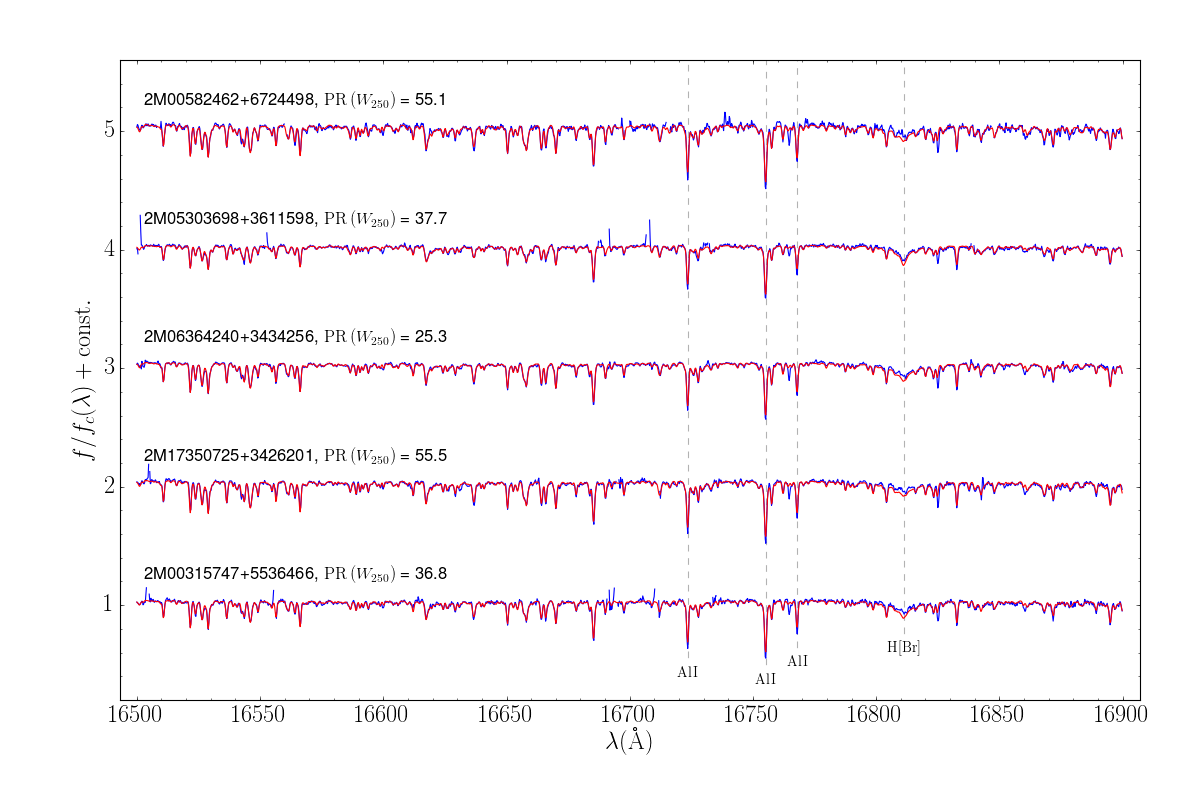} \label{fig:specs_most}}  \\ 
    
\subfloat[Three example pairs of spectroscopic twins. The twin spectra are over-plotted, one in green and the other in blue. Note that in the top example one of the stars lacks an ASPCAP fit, preventing identification as a twin via these parameters.]{%
	\includegraphics[width=\specfigwidth\columnwidth]{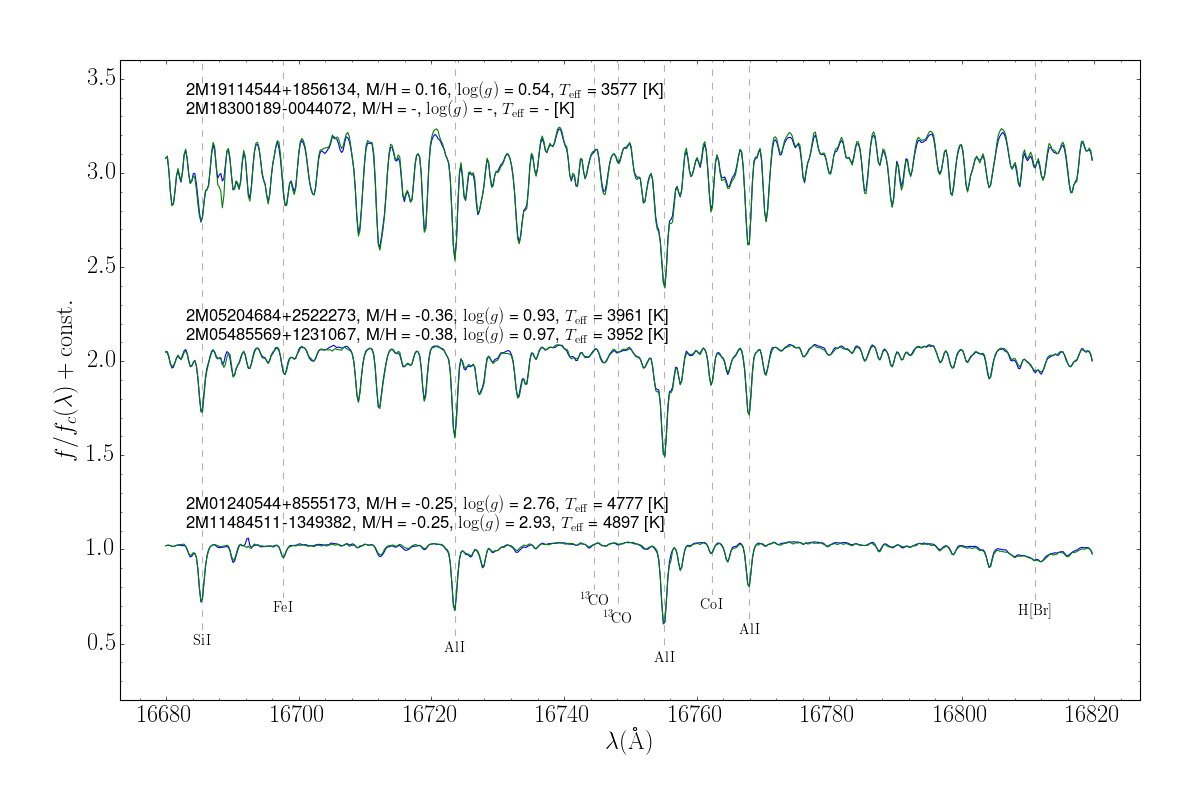} \label{fig:specs_twins}}
    
\caption{Example spectra for different groups of objects. The spectra plots were made using the APOGEE toolkit by \citet{bovy16}.  The red line is the ASPCAP fit and the blue line is the PCN spectrum, except where indicated. $\mathrm{PR}\left(W_{250}\right)$ indicates the percentile of objects with lower weirdness score. Relative fluxes are offset by a constant for display purposes. In every panel we choose the most informative wavelength range.}
 \label{fig:specs}
\end{figure*}

\begin{figure*}\ContinuedFloat
\centering
    \subfloat[Spectroscopic binaries.]{%
	\includegraphics[width=\specfigwidth\columnwidth]{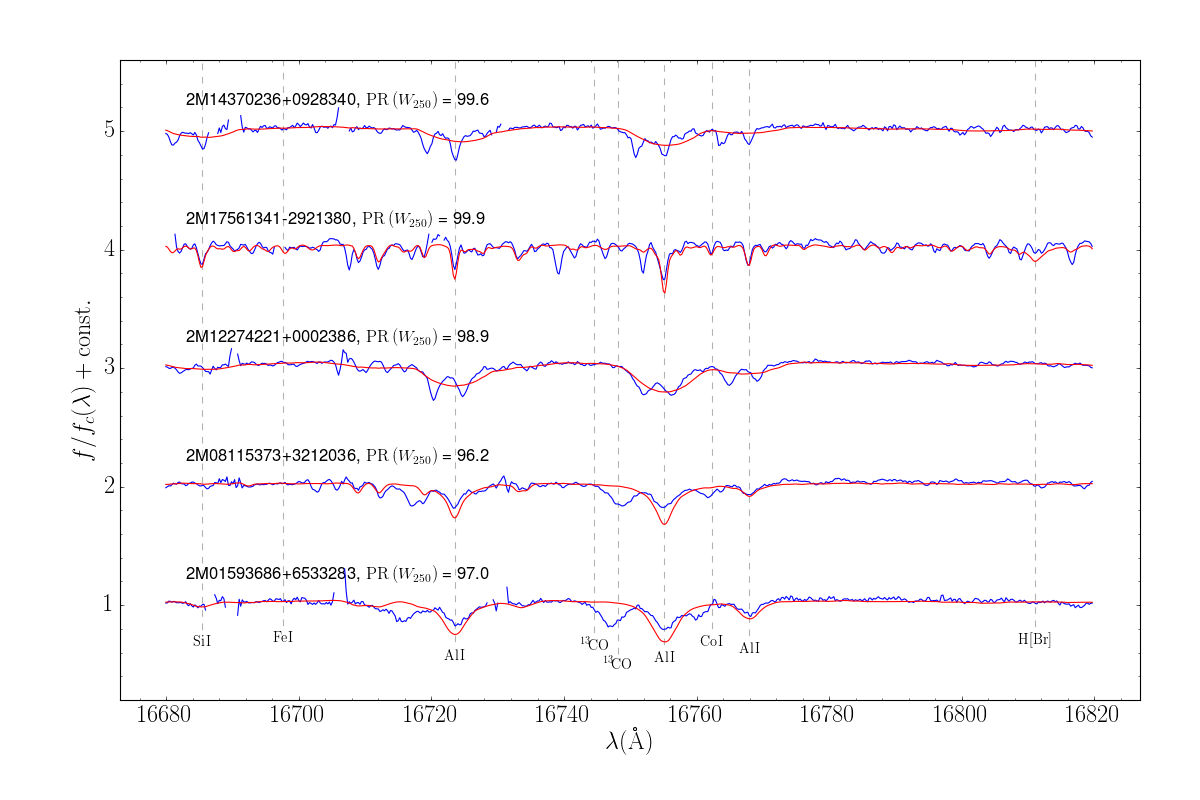} \label{fig:specs_sb2}}\\
    \subfloat[Fast rotators. Top spectrum is a typical APOGEE red giant, for comparison.]{%
	\includegraphics[width=\specfigwidth\columnwidth]{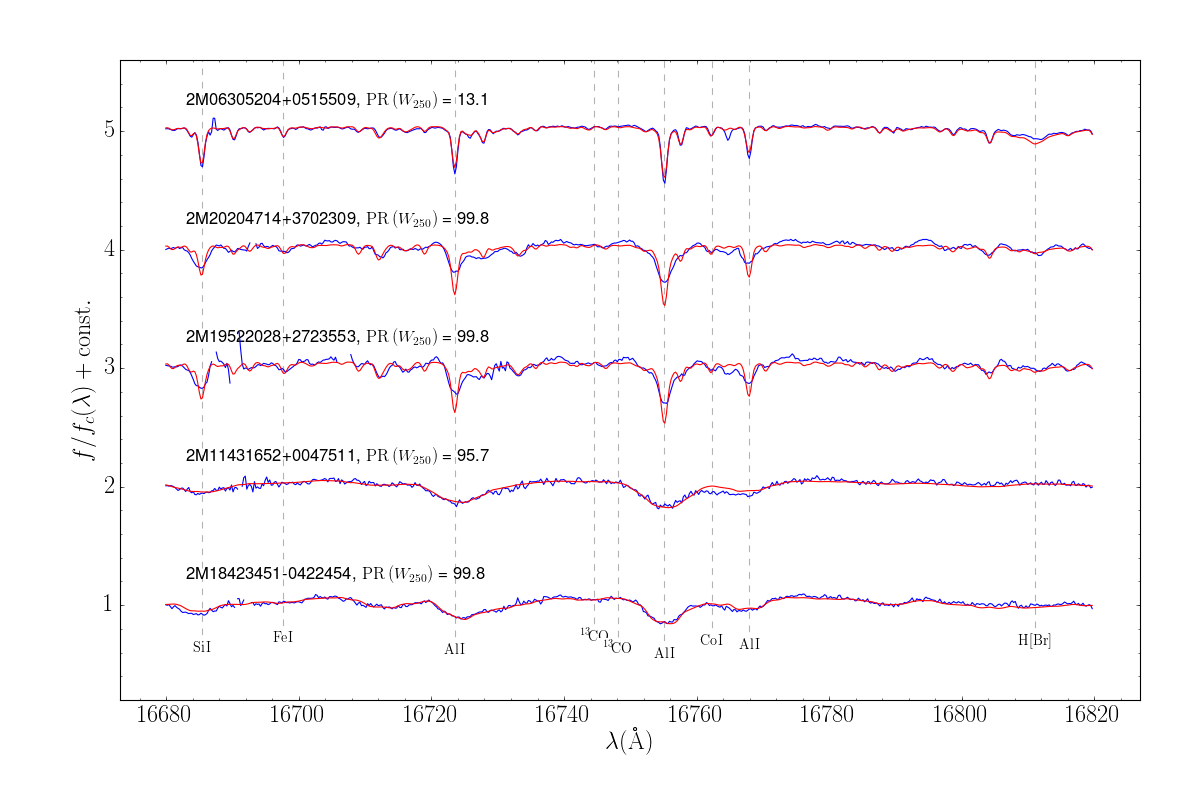}  \label{fig:specs_broad}}
\caption{Continued.}
\end{figure*}

\begin{figure*}\ContinuedFloat
\centering
    \subfloat[Carbon enriched stars. Top spectrum is a typical APOGEE red giant, for comparison.]{%
	\includegraphics[width=\specfigwidth\columnwidth]{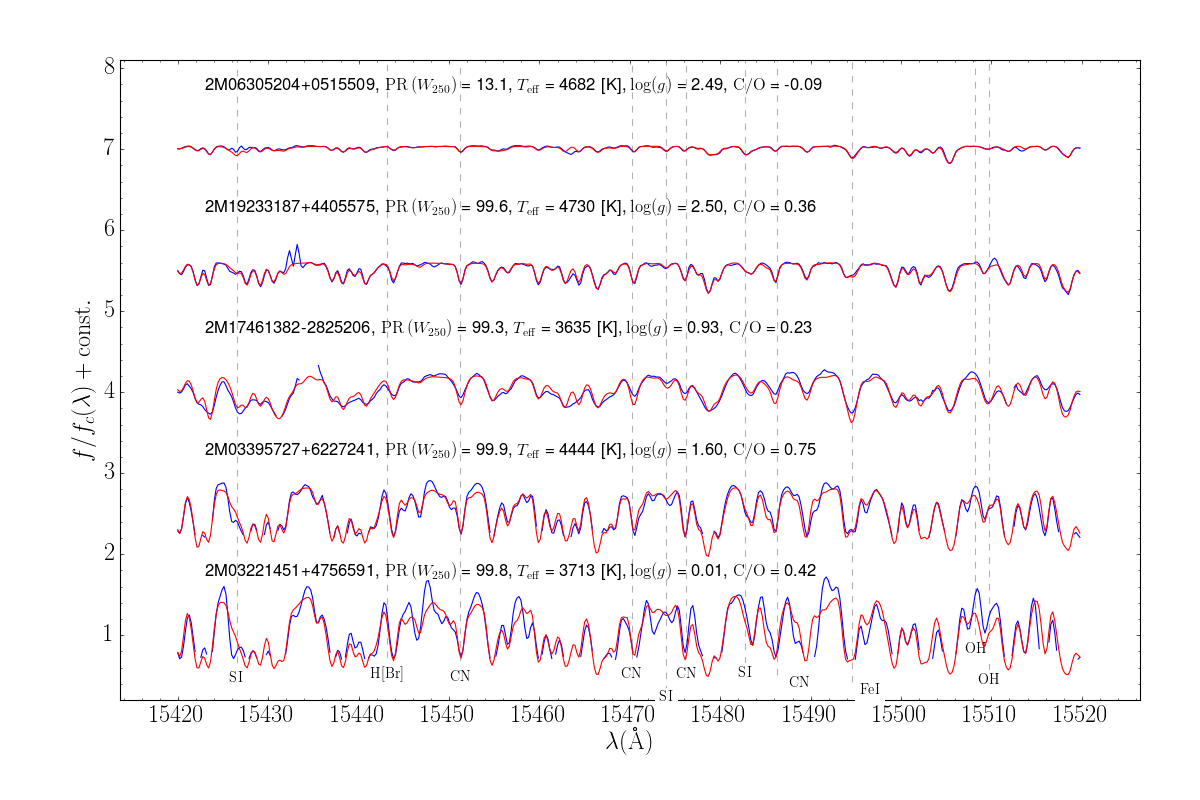} \label{fig:specs_carbon}}\\
    \subfloat[Stars with both absorption and hydrogen emission. Top spectrum is a typical APOGEE red giant, for comparison. We see both narrow and broad emission stars, and also both narrow and broad absorption. The second spectra from the top is also an SB2. The bottom spectrum has bad RV determination.]{%
    \includegraphics[width=\specfigwidth\columnwidth]{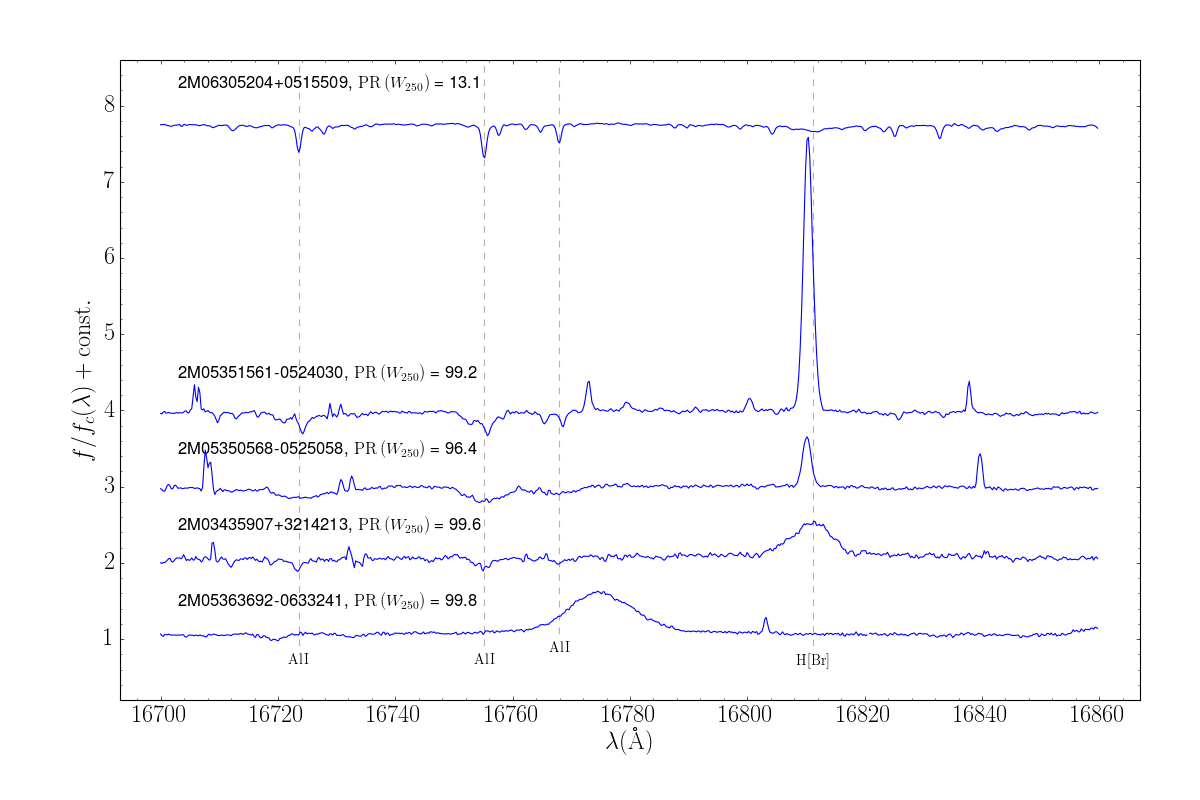} \label{fig:specs_em}} 
\caption{Continued.}
\end{figure*}

\begin{figure*}\ContinuedFloat
\centering

    \subfloat[M dwarfs. Top spectrum is a typical APOGEE red giant, for comparison. M dwarfs are detected as outliers due to their underrepresentation in the APOGEE dataset.]{%
	\includegraphics[width=\specfigwidth\columnwidth]{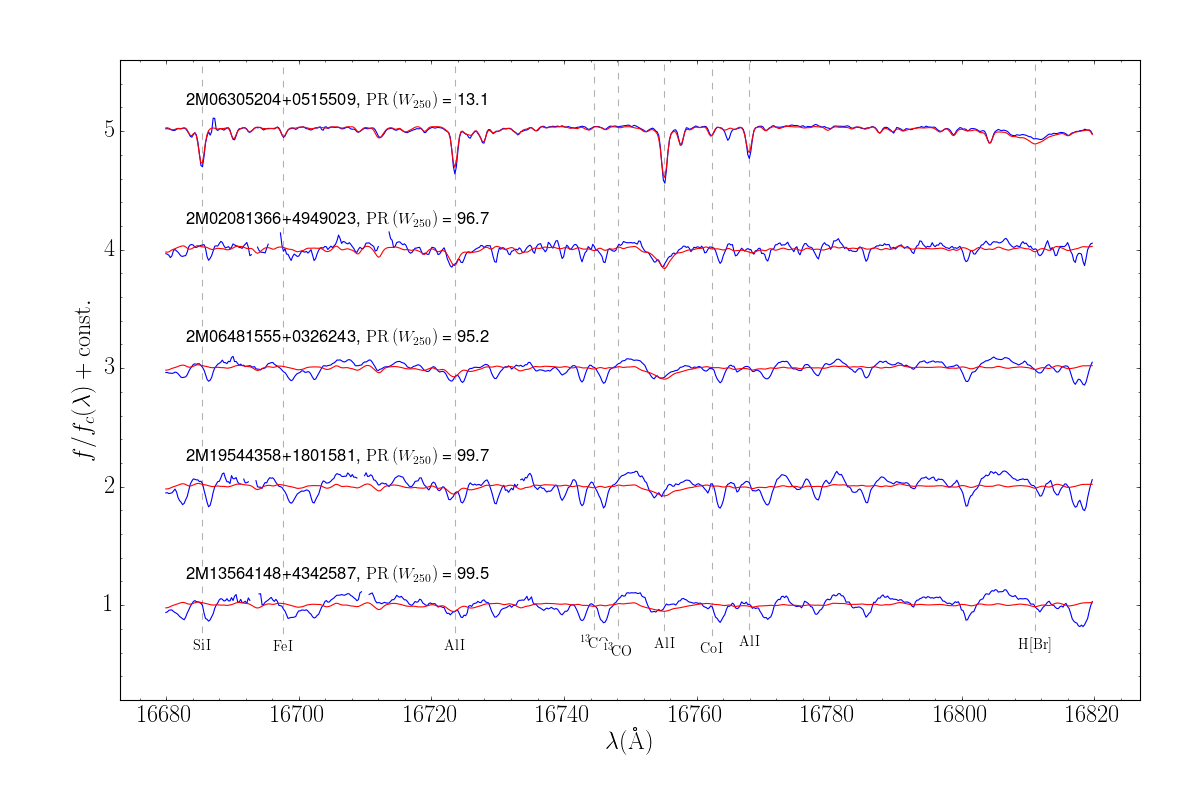} \label{fig:specs_mdwarf}}\\
    \subfloat[Stars from the 'others' pile. Top spectrum is a typical APOGEE red giant, for comparison. Starting from the second from top, the four outlying spectra are brown dwarf, massive star target, unexplained red giant, and a Wolf-Rayet star.]{%
    \includegraphics[width=\specfigwidth\columnwidth]{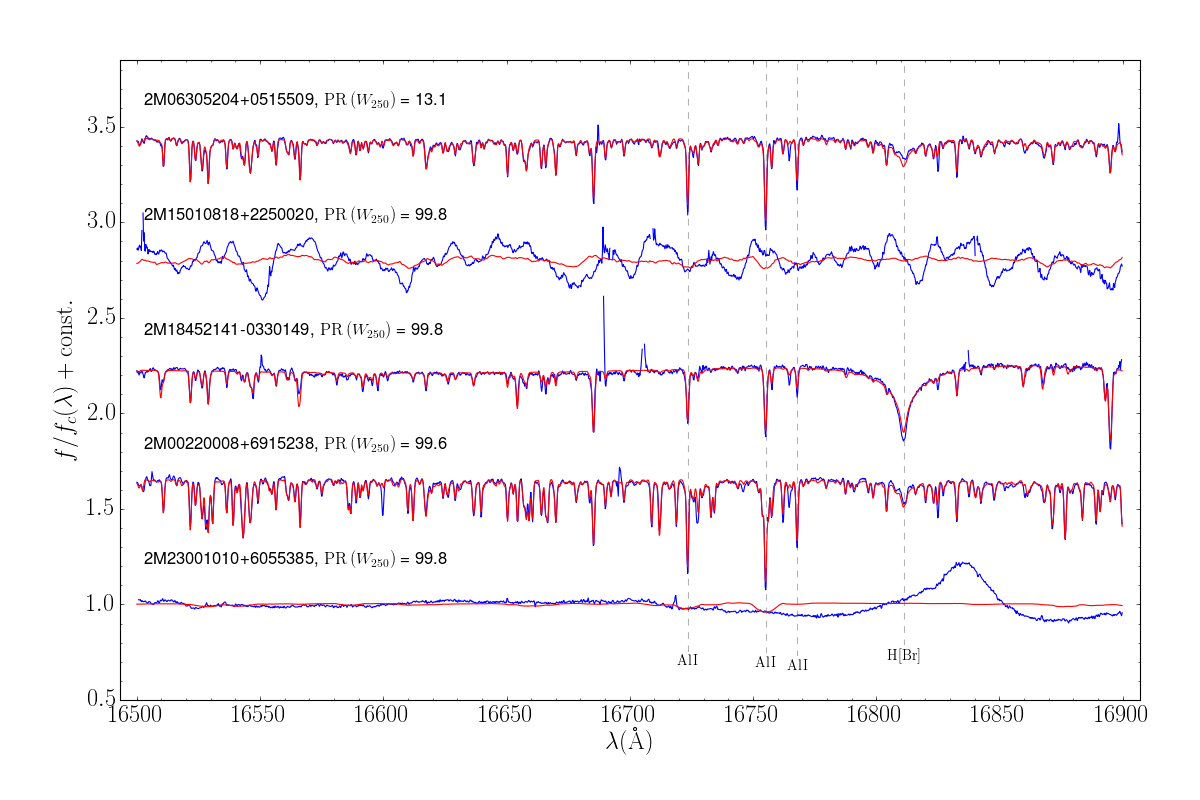} \label{fig:specs_other}}\\
\caption{Continued.}
\end{figure*}

\begin{figure*}\ContinuedFloat
\centering
    \subfloat[DR14 Faulty spectra. For the top two objects the problems are due to an issue in the PCN process, for the bottom three one of the visit spectra is bad.]{%
	\includegraphics[width=\specfigwidth\columnwidth]{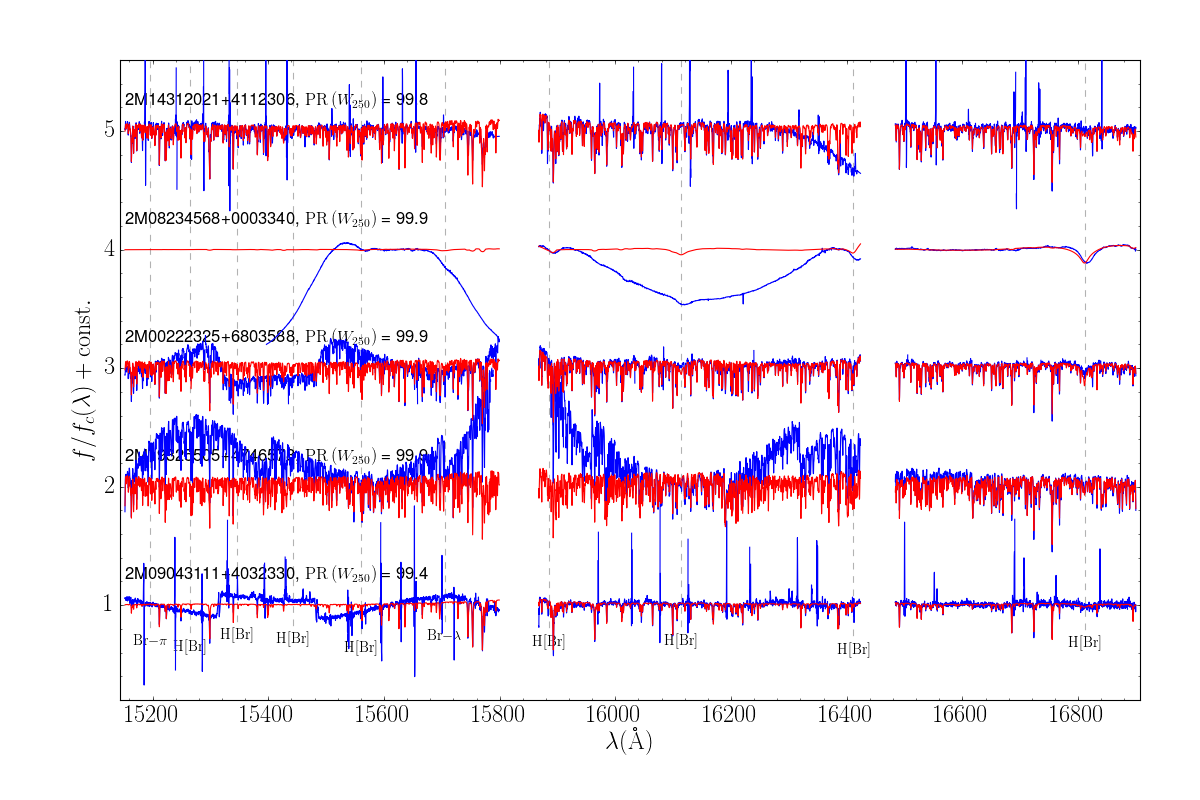} \label{fig:specs_bad}}\\
    
    \subfloat[B-type emission line stars showing double peaked hydrogen emission. The emission lines are not on the dotted lines due to wrong RV determination by the APOGEE pipeline.]{%
	\includegraphics[width=\specfigwidth\columnwidth]{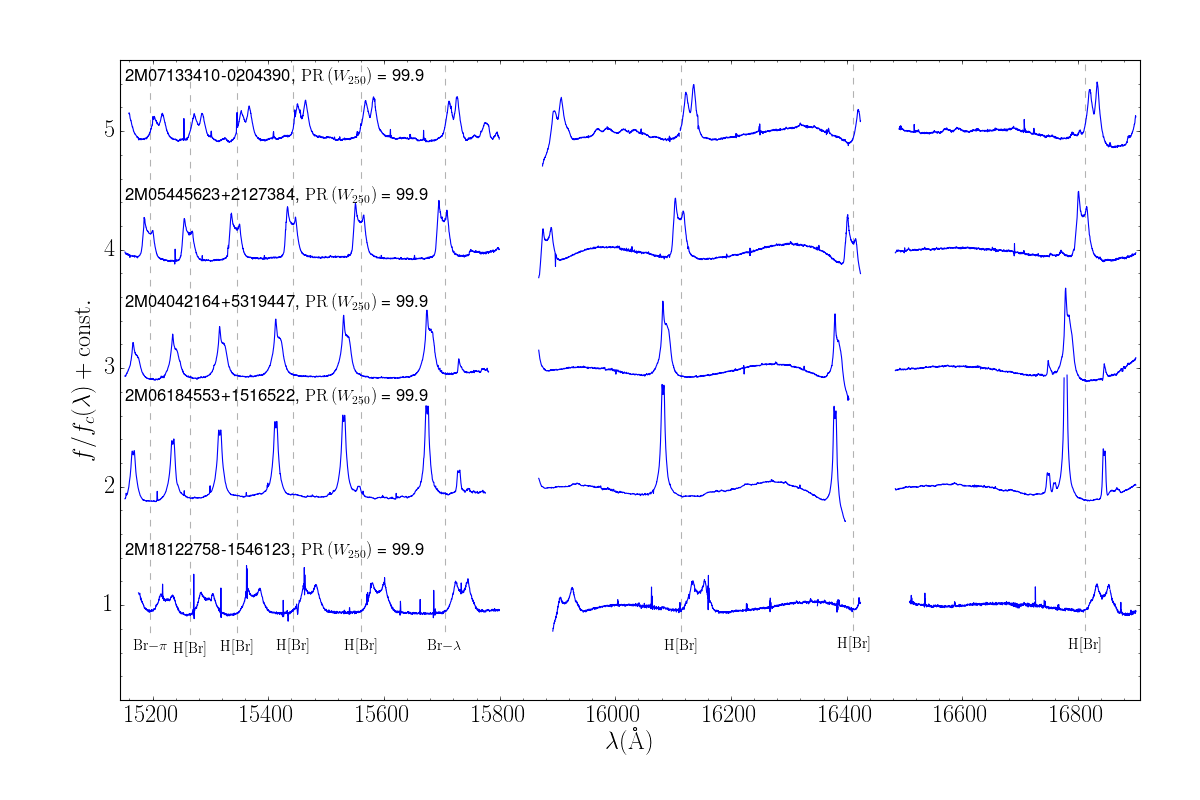} \label{fig:specs_be}} 
\caption{Continued.}
\end{figure*}

In Table \ref{tab:outliers_table} we present all the objects detected as outliers, and did not fall into any of the large groups. Tables with the objects in the rest of the groups are available online.

\begin{table*}
\centering
  \begin{tabular}{llll}
  \toprule
  APOGEE ID  &       RA [deg] &       DEC [deg] &                       Classification                                     \\
  \midrule
  2M15010818+2250020 &  225.284 &   22.8339 &                          Brown dwarf \\
  2M14323054+5049406 &  218.127 &    50.828 &                       Contact binary \\
  2M03114116-0043477 &  47.9215 &  -0.72993 &                       Contact binary \\
  2M14304297+0905087 &  217.679 &   9.08575 &                       Contact binary \\
  2M13465180+2257140 &  206.716 &   22.9539 &                       Contact binary \\
  2M14120965+0508201 &   213.04 &   5.13893 &                       Contact binary \\
  2M16145863+3016356 &  243.744 &   30.2766 &                       Contact binary \\
  2M03242324-0042148 &  51.0969 & -0.704119 &                       Contact binary \\
  2M03242324-0042148 &  51.0969 & -0.704119 &                       Contact binary \\
  2M16241043+4555265 &  246.043 &    45.924 &                       Contact binary \\
  2M16524137+4723275 &  253.172 &    47.391 &                       Contact binary \\
  2M06415063-0130177 &  100.461 &  -1.50494 &               Double-peaked emission \\
  2M04052624+5304494 &  61.3594 &   53.0804 &               Double-peaked emission \\
  2M13145725+1713303 &  198.739 &   17.2251 &                               Galaxy \\
  AP00425080+4117074 &  10.7117 &   41.2854 &                     Globular cluster \\
  AP00442956+4121359 &  11.1232 &     41.36 &                     Globular cluster \\
  AP00430957+4121321 &  10.7899 &   41.3589 &                     Globular cluster \\
  AP00424183+4051550 &  10.6743 &   40.8653 &                     Globular cluster \\
  AP00424183+4051550 &  10.6743 &   40.8653 &                     Globular cluster \\
  AP00431764+4127450 &  10.8235 &   41.4625 &                     Globular cluster \\
  2M18445087-0325251 &  281.212 &  -3.42364 &                         Massive star \\
  2M18452141-0330149 &  281.339 &  -3.50416 &                         Massive star \\
  2M18440079-0353160 &  281.003 &  -3.88778 &                         Massive star \\
  2M03411288+2453344 &  55.3037 &   24.8929 &                                   Unexplained non red giant \\
  2M05264478+1049152 &  81.6866 &   10.8209 &                                   Unexplained non red giant \\
  2M23375653+8534449 &  354.486 &   85.5792 &                                   Unexplained non red giant \\
  2M04255084+6007127 &  66.4619 &   60.1202 &                     Planetary nebula \\
  2M21021878+3641412 &  315.578 &   36.6948 &         Planetary nebula- Egg nebula \\
  2M18211606-1301256 &  275.317 &  -13.0238 &  Planetary nebula- Red Square nebula \\
  2M17534571-2949362 &   268.44 &  -29.8267 &                   Unexplained red giant \\
  2M06361326+0919120 &  99.0553 &   9.32001 &                  Unexplained red giant \\
  2M00220008+6915238 &  5.50037 &   69.2566 &                   Unexplained red giant \\
  2M21184119+4836167 &  319.672 &   48.6047 &                   Unexplained red giant \\
  2M20564714+5013372 &  314.196 &    50.227 &                   Unexplained red giant \\
  2M05501847-0010369 &   87.577 & -0.176939 &                   Unexplained red giant \\
  2M23001010+6055385 &  345.042 &   60.9274 &                      Wolf-Rayet star \\
  2M05473667+0020060 &  86.9028 &   0.33501 &                 Young stellar object \\
  \bottomrule
  \end{tabular}
  \caption{Outliers that and did not fall into any of the large groups.}
	\label{tab:outliers_table}
\end{table*}

In Table \ref{tab:new_be_stars} we list Be stars which are new in DR14 and thus not included in the \citet{chojnowski15} catalog. 

\begin{table*}
\centering
  \begin{tabular}{lll}
  \toprule
  APOGEE ID  &       RA [deg] &       DEC [deg]                                     \\
  \midrule
  2M20383016+2119439 &  309.626 &   21.3289 \\
  2M22425730+4443183 &  340.739 &   44.7218 \\
  2M06490825+0005220 &  102.284 &  0.089448 \\
  2M04480651+3359160 &  72.0271 &   33.9878 \\
  2M05284845+0209529 &  82.2019 &   2.16471 \\
  2M21582976+5429057 &  329.624 &   54.4849 \\
  2M22082542+5413262 &  332.106 &   54.2239 \\
  2M19322817-0454283 &  293.117 &  -4.90786 \\
  2M03145531+4841448 &  48.7305 &   48.6958 \\
  2M21523408+4713436 &  328.142 &   47.2288 \\
  2M04454937+4323302 &  71.4557 &   43.3917 \\
  2M18574904+1758251 &  284.454 &   17.9736 \\
  2M05312677+1101226 &  82.8616 &   11.0229 \\
  2M05384719-0235405 &  84.6967 &  -2.59459 \\
  2M22075623+5431064 &  331.984 &   54.5185 \\
  2M21380289+5037030 &  324.512 &   50.6175 \\
  2M06521036-0017440 &  103.043 &  -0.29556 \\
  2M04125427+6647203 &  63.2262 &    66.789 \\
  2M22142219+4206020 &  333.592 &   42.1006 \\
  2M04493134+3313091 &  72.3806 &   33.2192 \\
  2M02374876+5248458 &  39.4532 &   52.8127 \\
  2M05122466+4816538 &  78.1028 &   48.2816 \\
  2M23570808+6118272 &  359.284 &   61.3076 \\
  2M05441926+5241437 &  86.0803 &   52.6955 \\
  2M18042714-0958113 &  271.113 &  -9.96982 \\
  2M06514059+0019363 &  102.919 &  0.326773 \\
  2M02273460+4813548 &  36.8942 &   48.2319 \\
  2M23293672+4822513 &  352.403 &   48.3809 \\
  2M18040936-0827329 &  271.039 &  -8.45915 \\
  2M21151579+3235270 &  318.816 &   32.5909 \\
  2M06552851+2430188 &  103.869 &   24.5052 \\
  2M19575932+2714001 &  299.497 &   27.2334 \\
  2M04563331+6345566 &  74.1388 &   63.7657 \\
  2M19562230+2626258 &  299.093 &   26.4405 \\
  2M10214707+1532036 &  155.446 &   15.5344 \\
  2M05271779+1308569 &  81.8241 &   13.1492 \\
  2M02571539+4601118 &  44.3142 &     46.02 \\
  2M21504079+5518451 &   327.67 &   55.3125 \\
  2M04503901+3243187 &  72.6626 &   32.7219 \\
  2M22165865+6738450 &  334.244 &   67.6458 \\
  \bottomrule
  \end{tabular}
  \caption{Be stars.}
  \label{tab:new_be_stars}
\end{table*}

In Table \ref{tab:carbon} we list carbon rich stars that were detected as outliers.

\begin{table}
\centering
  \begin{tabular}{lll}
  \toprule
  APOGEE ID  &       RA [deg] &       DEC [deg]                                     \\
  \midrule
2M21095891+1111013 &  317.495 &  11.1837 \\
2M12553245+4328014 &  193.885 &  43.4671 \\
2M08031240+5311340 &  120.802 &  53.1928 \\
2M17552511-2517291 &  268.855 & -25.2914 \\
2M18442763-0614402 &  281.115 & -6.24452 \\
2M06211564-0124429 &  95.3152 & -1.41194 \\
2M12410240-0853066 &   190.26 & -8.88517 \\
2M13150364+1806426 &  198.765 &  18.1119 \\
2M07384226+2131021 &  114.676 &  21.5173 \\
2M05264861+2551545 &  81.7026 &  25.8652 \\
2M16334467-1343201 &  248.436 & -13.7223 \\
2M13381781-1458456 &  204.574 & -14.9793 \\
2M13122536+1313575 &  198.106 &  13.2327 \\
2M15000319+2955500 &  225.013 &  29.9306 \\
2M21330683+1209406 &  323.278 &  12.1613 \\
2M18455347-0328585 &  281.473 & -3.48293 \\
2M18495015-0235162 &  282.459 & -2.58786 \\
2M19425134+2235573 &  295.714 &  22.5993 \\
2M19474632+2349074 &  296.943 &  23.8187 \\
2M00242588+6221034 &  6.10785 &  62.3509 \\
2M04501927+3947587 &  72.5803 &  39.7996 \\
2M05012902+4023388 &  75.3709 &  40.3941 \\
2M21053099+2952201 &  316.379 &  29.8723 \\
2M01403590+6254392 &  25.1496 &  62.9109 \\
2M04405098+4705190 &  70.2124 &  47.0886 \\
2M18191371-1218145 &  274.807 &  -12.304 \\
2M18030503-2157460 &  270.771 & -21.9628 \\
2M18015024-2638220 &  270.459 & -26.6395 \\
2M17520031-2308488 &  268.001 & -23.1469 \\
2M18052874-2505351 &   271.37 & -25.0931 \\
2M18063056-2435442 &  271.627 & -24.5956 \\
2M18111704-2352577 &  272.821 & -23.8827 \\
2M18115753-1503100 &   272.99 & -15.0528 \\
2M18185547-1119080 &  274.731 & -11.3189 \\
2M18524968-2834454 &  283.207 & -28.5793 \\
2M17301939-2913292 &  262.581 & -29.2248 \\
2M19295061+0010102 &  292.461 &  0.16951 \\
2M19411240+4936344 &  295.302 &  49.6096 \\
2M19003459+4408290 &  285.144 &  44.1414 \\
2M19023427+4246148 &  285.643 &  42.7708 \\
2M19095794+4325272 &  287.491 &  43.4242 \\
2M06343313+0643006 &  98.6381 &  6.71686 \\
2M01471583+5753060 &   26.816 &   57.885 \\
  \bottomrule
  \end{tabular}
  \caption{Carbon rich stars.}
  \label{tab:carbon}
\end{table}

\begin{table}
\centering
  \begin{tabular}{lll}
  \toprule
  APOGEE ID  &       RA [deg] &       DEC [deg]                                     \\
  \midrule
2M21473632+5932259 &  326.901 &  59.5405 \\
2M21544864+5916346 &  328.703 &  59.2763 \\
2M11475977-0019182 &  176.999 & -0.32173 \\
2M18284700-1010553 &  277.196 &  -10.182 \\
2M17043371-2212322 &   256.14 &  -22.209 \\
2M19233187+4405575 &  290.883 &  44.0993 \\
2M12553245+4328014 &  193.885 &  43.4671 \\
2M14561660+1702441 &  224.069 &  17.0456 \\
2M15015733+2713595 &  225.489 &  27.2332 \\
2M15100330+3054073 &  227.514 &   30.902 \\
2M23290070+5711558 &  352.253 &  57.1989 \\
2M02403149+5600473 &  40.1312 &  56.0132 \\
2M07094794+0006382 &   107.45 &  0.11062 \\
2M07232483-0823577 &  110.853 & -8.39936 \\
2M03463234+3221127 &  56.6348 &  32.3536 \\
2M19531095+4635518 &  298.296 &  46.5977 \\
2M19200927+1317078 &  290.039 &  13.2855 \\
2M21315424+5219122 &  322.976 &  52.3201 \\
2M00334926+6837330 &  8.45529 &  68.6258 \\
2M04174731+4211335 &  64.4472 &  42.1926 \\
2M04195310+4109094 &  64.9713 &  41.1526 \\
2M06372981+0515011 &  99.3742 &  5.25033 \\
2M00373109+5743345 &  9.37956 &  57.7263 \\
2M03244820+6300289 &  51.2009 &   63.008 \\
2M03271166+6240211 &  51.7986 &  62.6725 \\
2M03330010+6330443 &  53.2504 &  63.5123 \\
2M03395727+6227241 &  54.9886 &  62.4567 \\
2M06393827+2403560 &  99.9095 &  24.0656 \\
2M05152962+2400147 &  78.8734 &  24.0041 \\
2M03103113+4831002 &  47.6297 &  48.5167 \\
2M03221451+4756591 &  50.5605 &  47.9498 \\
2M05522651+4329557 &  88.1105 &  43.4988 \\
2M20553607+5613011 &    313.9 &   56.217 \\
2M21031081+5414127 &  315.795 &  54.2369 \\
2M21084459+5442122 &  317.186 &  54.7034 \\
2M21590597+4539010 &  329.775 &  45.6503 \\
2M21554492+5414593 &  328.937 &  54.2498 \\
2M21573025+5440529 &  329.376 &  54.6814 \\
2M21594113+5351121 &  329.921 &  53.8534 \\
2M22085910+5434192 &  332.246 &   54.572 \\
2M18300408+0416050 &  277.517 &  4.26807 \\
2M04113023+2255071 &   62.876 &  22.9187 \\
2M06531594-0439506 &  103.316 & -4.66407 \\
  \bottomrule
  \end{tabular}
\end{table}

In Table \ref{tab:binaries} we list the spectroscopic binaries that were detected as outliers.

\begin{table}
\centering
  \begin{tabular}{lll}
  \toprule
  APOGEE ID  &       RA [deg] &       DEC [deg]                                     \\
  \midrule
2M14251536+3915337 &  216.314 &   39.2594 \\
2M08115373+3212036 &  122.974 &    32.201 \\
2M12274221+0002386 &  186.926 &  0.044058 \\
2M05284223+4359528 &   82.176 &    43.998 \\
2M05240837+2711064 &  81.0349 &   27.1851 \\
2M03563567+7857072 &  59.1487 &    78.952 \\
2M09314691+5618248 &  142.945 &   56.3069 \\
2M13413548-1723167 &  205.398 &   -17.388 \\
2M01193634+8435481 &  19.9014 &   84.5967 \\
2M14542303+3122323 &  223.596 &   31.3756 \\
2M09315645+3714213 &  142.985 &   37.2393 \\
2M10280514+1735219 &  157.021 &   17.5894 \\
2M11081296-1205110 &  167.054 &  -12.0864 \\
2M21302403+1132483 &    322.6 &   11.5468 \\
2M15002128+3645004 &  225.089 &   36.7501 \\
2M18460678-0337057 &  281.528 &  -3.61827 \\
2M19430973+2357587 &  295.791 &   23.9663 \\
2M19225746+3824509 &  290.739 &   38.4141 \\
2M21523747+3853140 &  328.156 &   38.8872 \\
2M18054943-3059442 &  271.456 &  -30.9956 \\
2M18075069-3116452 &  271.961 &  -31.2792 \\
2M18081808-2553287 &  272.075 &  -25.8913 \\
2M17561341-2921380 &  269.056 &  -29.3606 \\
2M17360668-2710099 &  264.028 &  -27.1694 \\
2M18103554-1811011 &  272.648 &  -18.1836 \\
2M18192203-1411326 &  274.842 &  -14.1924 \\
2M18192899-1452043 &  274.871 &  -14.8679 \\
2M18280206-1217422 &  277.009 &  -12.2951 \\
2M17345651-2048568 &  263.735 &  -20.8158 \\
2M18001201-2631398 &   270.05 &  -26.5277 \\
2M18041435-2455385 &   271.06 &  -24.9274 \\
2M18165573-1852394 &  274.232 &  -18.8776 \\
2M18040248-1805575 &   271.01 &  -18.0993 \\
2M17464152-2713191 &  266.673 &   -27.222 \\
2M17531813-2816161 &  268.326 &  -28.2711 \\
2M18042203-2917298 &  271.092 &  -29.2916 \\
2M17282574-2906578 &  262.107 &  -29.1161 \\
2M17285197-2815064 &  262.217 &  -28.2518 \\
2M18104783-2824046 &  272.699 &  -28.4013 \\
2M19383737+4957227 &  294.656 &   49.9563 \\
2M19454606+5113275 &  296.442 &   51.2243 \\
2M19301580+4932086 &  292.566 &   49.5357 \\
2M18544916+4512355 &  283.705 &   45.2099 \\
2M19123630+4603326 &  288.151 &   46.0591 \\
2M01593686+6533283 &  29.9036 &   65.5579 \\
2M14370236+0928340 &   219.26 &   9.47612 \\
  \bottomrule
  \end{tabular}
  \caption{Spectroscopic binaries.}
  \label{tab:binaries}
\end{table}

\begin{table}
\centering
  \begin{tabular}{lll}
  \toprule
  APOGEE ID  &       RA [deg] &       DEC [deg]                                     \\
  \midrule
2M15021575+2319460 &  225.566 &   23.3295 \\
2M11254661+5217235 &  171.444 &   52.2899 \\
2M11012916+1215329 &  165.372 &   12.2592 \\
2M13405651+0031563 &  205.235 &  0.532321 \\
2M18411589-1016542 &  280.316 &  -10.2817 \\
2M19564877+4458058 &  299.203 &   44.9683 \\
2M20034832+4536148 &  300.951 &   45.6041 \\
2M19190180+4153127 &  289.758 &   41.8869 \\
2M19561994+4120265 &  299.083 &   41.3407 \\
2M13483079+1750445 &  207.128 &   17.8457 \\
2M12115853+1425463 &  182.994 &   14.4295 \\
2M12462044+1251325 &  191.585 &   12.8591 \\
2M12505092+1324147 &  192.712 &   13.4041 \\
2M14123798+5426481 &  213.158 &   54.4467 \\
2M11542519+5554150 &  178.605 &   55.9042 \\
2M11044917+4840467 &  166.205 &   48.6797 \\
2M14232001+0541233 &  215.833 &   5.68982 \\
2M16582628+0939165 &   254.61 &   9.65459 \\
2M09242547-0650183 &  141.106 &  -6.83842 \\
2M19400944+3832454 &  295.039 &    38.546 \\
2M00065508+0154022 &  1.72953 &   1.90061 \\
2M05502340+0420349 &  87.5975 &   4.34304 \\
2M07054011+3812529 &  106.417 &   38.2147 \\
2M07250686+2435451 &  111.279 &   24.5959 \\
2M05345563-0601036 &  83.7318 &  -6.01768 \\
2M05350392-0529033 &  83.7663 &  -5.48426 \\
2M05360185-0517365 &  84.0077 &  -5.29349 \\
2M05350138-0615175 &  83.7558 &  -6.25487 \\
2M05351236-0543184 &  83.8015 &   -5.7218 \\
2M05351561-0524030 &  83.8151 &  -5.40085 \\
2M05351798-0604430 &  83.8249 &  -6.07862 \\
2M05371161-0723239 &  84.2984 &  -7.38999 \\
2M19383668+4723194 &  294.653 &   47.3887 \\
2M18534305+0026394 &  283.429 &  0.444304 \\
2M04135110+4938317 &   63.463 &   49.6422 \\
2M03361242+4651208 &  54.0518 &   46.8558 \\
2M17393731-2324309 &  264.905 &  -23.4086 \\
2M17340500-2808243 &  263.521 &  -28.1401 \\
2M18234612-1501159 &  275.942 &  -15.0211 \\
2M17190649-2745172 &  259.777 &  -27.7548 \\
2M17380171-2858281 &  264.507 &  -28.9745 \\
2M17535762-2841520 &   268.49 &  -28.6978 \\
2M17144370-2449231 &  258.682 &  -24.8231 \\
2M17364991-2728343 &  264.208 &  -27.4762 \\
2M19252567+4229371 &  291.357 &   42.4936 \\
  \bottomrule
  \end{tabular}
\end{table}

In Table \ref{tab:broad} we list the spectroscopic binaries that were detected as outliers.

\begin{table*}
\centering
  \begin{tabular}{lll}
  \toprule
  APOGEE ID  &       RA [deg] &       DEC [deg]                                     \\
  \midrule
2M21031344+0942207 &  315.806 &   9.70577 \\
2M07365631+4517467 &  114.235 &   45.2963 \\
2M07560603+2626563 &  119.025 &    26.449 \\
2M17550303-2557141 &  268.763 &  -25.9539 \\
2M18423451-0422454 &  280.644 &   -4.3793 \\
2M11431652+0047511 &  175.819 &  0.797531 \\
2M04131296+5546540 &   63.304 &   55.7817 \\
2M03283689+7947391 &  52.1537 &   79.7942 \\
2M16132421+5140269 &  243.351 &   51.6742 \\
2M13553588+4436441 &    208.9 &   44.6123 \\
2M18451898-0150567 &  281.329 &  -1.84909 \\
2M19421896+2426209 &  295.579 &   24.4392 \\
2M21131747+4843554 &  318.323 &   48.7321 \\
2M20111813+2058271 &  302.826 &   20.9742 \\
2M03464878+2304074 &  56.7033 &   23.0687 \\
2M19142629+1202560 &   288.61 &   12.0489 \\
2M20204714+3702309 &  305.196 &   37.0419 \\
2M19014937+0520105 &  285.456 &   5.33626 \\
2M03220356+5654161 &  50.5148 &   56.9045 \\
2M17434496-2941008 &  265.937 &  -29.6836 \\
2M18142425-1911037 &  273.601 &  -19.1844 \\
2M18202527-1537239 &  275.105 &  -15.6233 \\
2M18313707-1222341 &  277.904 &  -12.3761 \\
2M19154842+4636261 &  288.952 &   46.6073 \\
2M19544569+4041406 &   298.69 &   40.6946 \\
2M20000263+4529265 &  300.011 &   45.4907 \\
2M18543899+0012432 &  283.662 &  0.212021 \\
2M19522028+2723553 &  298.085 &   27.3987 \\
  \bottomrule
  \end{tabular}
  \caption{Fast rotators.}
  \label{tab:broad}
\end{table*}

In Table \ref{tab:bad_reduct} we list the objects with bad DR14 reductions.

\begin{table}
\centering
  \begin{tabular}{lll}
  \toprule
  APOGEE ID  &       RA [deg] &       DEC [deg]                                     \\
  \midrule
2M00354276+8619045 &  8.92818 &   86.3179 \\
2M19315445+4813349 &  292.977 &   48.2264 \\
2M19294950+4740246 &  292.456 &   47.6735 \\
2M00250046+5503033 &  6.25194 &   55.0509 \\
2M19205656+4846274 &  290.236 &   48.7743 \\
2M19410822+4019319 &  295.284 &   40.3255 \\
2M19325505+4746578 &  293.229 &   47.7827 \\
2M19432504+2229419 &  295.854 &    22.495 \\
2M06365780+0702069 &  99.2409 &   7.03526 \\
2M19100818-0553311 &  287.534 &  -5.89197 \\
2M07542422+3916064 &  118.601 &   39.2685 \\
2M19193061+4842214 &  289.878 &    48.706 \\
2M19341894+4800216 &  293.579 &    48.006 \\
2M18315699-0100106 &  277.987 &  -1.00296 \\
2M21223490+5110033 &  320.645 &   51.1676 \\
2M14315024+5101159 &  217.959 &   51.0211 \\
2M03292627+4656162 &  52.3595 &   46.9379 \\
2M05322756+2658537 &  83.1149 &   26.9816 \\
2M19130107-0549328 &  288.254 &  -5.82579 \\
2M19411184+4013301 &  295.299 &    40.225 \\
2M19455347+2412201 &  296.473 &   24.2056 \\
2M14283924+4014496 &  217.164 &   40.2471 \\
2M19570041+2059538 &  299.252 &   20.9983 \\
2M19522176+1840186 &  298.091 &   18.6718 \\
2M20464928+3411241 &  311.705 &     34.19 \\
2M14273401+4014470 &  216.892 &   40.2464 \\
2M20353553+5428403 &  308.898 &   54.4779 \\
2M19343359+4823093 &   293.64 &   48.3859 \\
2M03324489+4623388 &  53.1871 &   46.3941 \\
2M19441693+4905154 &  296.071 &   49.0876 \\
2M07014143+0449051 &  105.423 &   4.81809 \\
2M21201614-0109393 &  320.067 &  -1.16092 \\
2M08235914+0008354 &  125.996 &  0.143176 \\
2M23583343+5635047 &  359.639 &   56.5847 \\
2M23230618+5733020 &  350.776 &   57.5506 \\
2M06381497+0557479 &  99.5624 &   5.96331 \\
2M17470159-2849173 &  266.757 &  -28.8215 \\
2M04424759+3825359 &  70.6983 &   38.4267 \\
2M19095216+1120219 &  287.467 &   11.3394 \\
2M17103385+3641103 &  257.641 &   36.6862 \\
2M17192832+5804145 &  259.868 &   58.0707 \\
2M07591385+4049311 &  119.808 &   40.8253 \\
2M08013264+4307298 &  120.386 &    43.125 \\
2M09095175+4254040 &  137.466 &   42.9011 \\
2M09104765+4139238 &  137.699 &   41.6566 \\
2M10265734+4149117 &  156.739 &   41.8199 \\
2M10400281+4306255 &  160.012 &   43.1071 \\
2M16011348+4149493 &  240.306 &   41.8304 \\
2M16023049+3949503 &  240.627 &   39.8306 \\
2M16034776+4051552 &  240.949 &   40.8653 \\
2M16034893+4047314 &  240.954 &   40.7921 \\
2M16042629+4030585 &   241.11 &   40.5163 \\
2M16060267+4042385 &  241.511 &   40.7107 \\
2M16062342+4023224 &  241.598 &   40.3896 \\
2M16065762+4012407 &   241.74 &   40.2113 \\
2M16103330+4146123 &  242.639 &   41.7701 \\
2M16111200+4132006 &    242.8 &   41.5335 \\
2M14250643+3912427 &  216.277 &   39.2119 \\
2M14263122+3921276 &   216.63 &   39.3577 \\
2M14264018+4018477 &  216.667 &   40.3133 \\
2M14270892+4008013 &  216.787 &   40.1337 \\
2M14285271+4015518 &   217.22 &   40.2644 \\
  \bottomrule
  \end{tabular}
  \caption{Bad reductions.}
  \label{tab:bad_reduct}
\end{table}

\begin{table}
\centering
  \begin{tabular}{lll}
  \toprule
  APOGEE ID  &       RA [deg] &       DEC [deg]                                     \\
  \midrule
2M06285236+0007407 &  97.2182 &  0.127981 \\
2M13441054+2735078 &  206.044 &   27.5855 \\
2M16280255-1306104 &  247.011 &  -13.1029 \\
2M17181861+4206399 &  259.578 &   42.1111 \\
2M09082892+3618428 &  137.121 &   36.3119 \\
2M13044971+7301298 &  196.207 &    73.025 \\
2M09294773+5544429 &  142.449 &   55.7453 \\
2M12242677+2534571 &  186.112 &   25.5826 \\
2M12283815+2613370 &  187.159 &    26.227 \\
2M12284457+2553575 &  187.186 &   25.8993 \\
2M10265302+1713099 &  156.721 &   17.2194 \\
2M10282637+1545209 &   157.11 &   15.7558 \\
2M21310488+1250496 &   322.77 &   12.8471 \\
2M13500810+4233262 &  207.534 &   42.5573 \\
2M10521368+0101300 &  163.057 &   1.02502 \\
2M21103095+4741321 &  317.629 &   47.6923 \\
2M18312899-0138055 &  277.871 &  -1.63487 \\
2M18350820+0002348 &  278.784 &  0.043008 \\
2M20510547+5125023 &  312.773 &   51.4173 \\
2M20535239+4932004 &  313.468 &   49.5334 \\
2M21302584+4452299 &  322.608 &    44.875 \\
2M20321595+5337112 &  308.066 &   53.6198 \\
2M07115139+0539169 &  107.964 &   5.65472 \\
2M18043735+0155085 &  271.156 &   1.91904 \\
2M20412525+3317111 &  310.355 &   33.2864 \\
2M19532259+0424013 &  298.344 &   4.40037 \\
2M04291231+3515567 &  67.3013 &   35.2658 \\
2M16553254-2134100 &  253.886 &  -21.5695 \\
2M17564183-2803554 &  269.174 &  -28.0654 \\
2M19281906+4915086 &  292.079 &   49.2524 \\
2M19331420+4841507 &  293.309 &   48.6974 \\
2M19343984+4809524 &  293.666 &   48.1646 \\
2M19094607+3747391 &  287.442 &   37.7942 \\
2M23483899+6452355 &  357.162 &   64.8765 \\
2M04060214+4655320 &  61.5089 &   46.9256 \\
2M22190955-0133473 &   334.79 &  -1.56315 \\
2M22472985+0553172 &  341.874 &   5.88812 \\
2M14442821+4511096 &  221.118 &    45.186 \\
2M14493515+4634280 &  222.396 &   46.5745 \\
2M14140537+5438148 &  213.522 &   54.6375 \\
2M23240792+5732077 &  351.033 &   57.5355 \\
2M05060092+3556109 &  76.5038 &   35.9364 \\
2M06202991+0723133 &  95.1246 &   7.38705 \\
2M03353783+3140491 &  53.9077 &   31.6803 \\
2M05350478-0443546 &  83.7699 &  -4.73184 \\
2M21425212+6955149 &  325.717 &   69.9208 \\
2M00474266+0351290 &  11.9278 &   3.85806 \\
2M02275302-0855544 &   36.971 &   -8.9318 \\
2M02310705-0758192 &  37.7794 &  -7.97201 \\
2M02325195-0806163 &  38.2165 &  -8.10455 \\
2M02332095-0903458 &  38.3373 &  -9.06273 \\
2M14484064-0706253 &  222.169 &  -7.10704 \\
2M14362130+5733384 &  219.089 &   57.5607 \\
2M06391017+0518525 &  99.7924 &    5.3146 \\
2M09235450+2753539 &  140.977 &   27.8983 \\
2M09260229+2839009 &   141.51 &   28.6503 \\
2M00525338+3832558 &  13.2224 &   38.5488 \\
2M15122530+6658305 &  228.105 &   66.9752 \\
2M05495923+4136264 &  87.4968 &   41.6073 \\
2M06232278-0441150 &   95.845 &  -4.68751 \\
2M11540771+1810106 &  178.532 &   18.1696 \\
2M15044648+2224548 &  226.194 &   22.4152 \\
  \bottomrule
  \end{tabular}
\end{table}


\bsp	
\label{lastpage}
\end{document}